\documentclass[aps,prd,preprintnumbers,groupedaddress,nofootinbib,amssymb,notitlepage,eqsecnum]{revtex4-2}
\usepackage{here}
\usepackage[dvipdfmx]{graphicx}
\usepackage{amsmath,amsthm,amssymb}
\usepackage{bm}
\usepackage{color}

\usepackage{amsfonts}
\usepackage{dcolumn}
\usepackage{ulem}
\usepackage{hyperref}
\allowdisplaybreaks[1]
\usepackage{stackengine}

\newcommand{\be}{\begin{equation}}  
\newcommand{\ee}{\end{equation}}
\newcommand{\ba}{\begin{eqnarray}}
\newcommand{\ea}{\end{eqnarray}}

\newcommand{\rd}{{\rm d}}

\newcommand{\bem}{\begin{bmatrix}}
\newcommand{\eem}{\end{bmatrix}}
\newcommand{\Mpl}{M_{\rm Pl}}

\allowdisplaybreaks

\begin{document}

\preprint{RIKEN-iTHEMS-Report-24, YITP-24-45, WUCG-24-04}

\title{Linear stability of vector Horndeski black holes}

\author{Che-Yu Chen$^{1}$\footnote{{\tt b97202056@gmail.com}}}
\author{Antonio De Felice$^{2}$\footnote{{\tt antonio.defelice@yukawa.kyoto-u.ac.jp}}}
\author{Shinji Tsujikawa$^{3}$\footnote{{\tt tsujikawa@waseda.jp}}}

\affiliation{$^1$RIKEN iTHEMS, Wako, Saitama 351-0198, Japan\\
$^2$Center for Gravitational Physics and Quantum Information, 
Yukawa Institute for Theoretical Physics, Kyoto University, 
606-8502, Kyoto, Japan\\
$^3$Department of Physics, Waseda University, 
3-4-1 Okubo, Shinjuku, Tokyo 169-8555, Japan}

\begin{abstract}

Horndeski's vector-tensor (HVT) gravity is described by 
a Lagrangian in which the field strength 
$F_{\mu \nu}=\partial_{\mu}A_{\nu}-\partial_{\nu}A_{\mu}$ 
of a vector field $A_{\mu}$ interacts with a double dual 
Riemann tensor $L^{\mu\nu\alpha\beta}$ in the form 
$\beta L^{\mu\nu\alpha\beta}F_{\mu \nu} F_{\alpha \beta}$, 
where $\beta$ is a constant. 
In Einstein-Maxwell-HVT theory, there are static and 
spherically symmetric black hole (BH) solutions with 
electric or magnetic charges, whose metric components 
are modified from those in the Reissner-Nordstr\"om geometry. The electric-magnetic duality of solutions is broken even at the background level by the nonvanishing coupling constant $\beta$. 
We compute a second-order action of BH perturbations 
containing both the odd- and even-parity modes and 
show that there are four dynamical perturbations
arising from the gravitational and vector-field sectors. 
We derive all the linear stability conditions associated 
with the absence of ghosts and radial/angular Laplacian 
instabilities for both the electric and magnetic BHs.
These conditions exhibit the difference between the
electrically and magnetically charged cases 
by reflecting the breaking of electric-magnetic duality 
at the level of perturbations. 
In particular, the four angular propagation speeds 
in the large-multipole limit are different from each 
other for both the electric and magnetic BHs. 
This suggests the breaking of eikonal correspondence 
between the peak position of at least one of the potentials of dynamical perturbations and the radius 
of photon sphere. 
For the electrically and magnetically charged cases, 
we elucidate parameter spaces of the HVT coupling and the BH charge in which the BHs without naked singularities are 
linearly stable.

\end{abstract}

\date{\today}


\maketitle

\section{Introduction}
\label{introsec}

The physics of black holes (BHs) can now be probed 
by the observations of gravitational waves \cite{LIGOScientific:2016aoc,LIGOScientific:2018mvr,LIGOScientific:2020tif} as well as BH shadows \cite{EventHorizonTelescope:2019dse}. 
General Relativity (GR) is a fundamental theory for describing the gravitational interaction in both strong and weak gravity regimes. While the accuracy of GR has been well-tested by the 
solar-system experiments \cite{Will:2014kxa} and 
submillimeter laboratory tests \cite{Hoyle:2000cv,Adelberger:2003zx},
we cannot exclude the possibility of deviation from GR 
on highly curved backgrounds. 
On the cosmological side, the long-standing problems of 
dark matter and dark energy may suggest the existence of 
propagating degrees of freedom (DOFs) beyond those appearing 
in standard model of particle physics 
and GR \cite{Bertone:2004pz,Copeland:2006wr}.
If such new DOFs also manifest themselves in strong 
gravity regimes, it is possible to probe their signatures from the observations of gravitational waves and BH shadows \cite{Berti:2015itd,Barack:2018yly,Berti:2018cxi,Berti:2018vdi}.

If there is a scalar field $\phi$ directly coupled 
to gravity, the metric on a static and spherically symmetric background 
can be modified from the Schwarzschild BH solution. 
In the presence of a nonminimal coupling of the form $F(\phi)R$, 
where $R$ is a Ricci scalar, it is known that the BH does not 
acquire an additional scalar hair \cite{Hawking:1972qk,Bekenstein:1995un,Sotiriou:2011dz,Hui:2012qt}. 
This is attributed to the fact that $R=0$ on a vacuum 
Schwarzschild background.
We can think of the other scalar-gravitational coupling 
of the form $F(\phi)R_{\rm GB}^2$, where $R_{\rm GB}^2$ 
is a Gauss-Bonnet curvature invariant. 
Since the GB invariant is nonvanishing on the
vacuum Schwarzschild background, it is possible 
to realize nontrivial BH solutions 
with scalar hairs (see e.g., Refs.~\cite{Kanti:1995vq,Torii:1996yi,Kanti:1997br,Sotiriou:2013qea,Doneva:2017bvd,Silva:2017uqg,Antoniou:2017acq,Minamitsuji:2018xde}).
We note that the scalar-Gauss-Bonnet coupling belongs to 
a sub-class of Horndeski theories
with second-order field equations 
of motion \cite{Horndeski:1974wa,Deffayet:2011gz,Kobayashi:2011nu,Charmousis:2011bf}. 
Even in full Horndeski theories, the scalar-Gauss-Bonnet coupling plays a prominent role in the existence of asymptotically flat, 
linearly stable, hairy BH solutions \cite{Minamitsuji:2022mlv,Minamitsuji:2022vbi}.

Instead of the scalar field, we can consider a vector 
field $A_{\mu}$ coupled to gravity. 
In 1976, Horndeski constructed $U(1)$ 
gauge-invariant vector-tensor theories with 
second-order field equations of motion \cite{Horndeski:1976gi}. 
The second-order property of field equations can 
prevent the appearance of Ostrogradsky ghosts associated with higher-order derivative terms.
In this case, there is a unique vector-tensor interaction 
of the form ${\cal L}_{\rm HVT}=
\beta L^{\mu \nu \alpha \beta} F_{\mu \nu}F_{\alpha \beta}$, where $\beta$ is a coupling constant, 
$L^{\mu \nu \alpha \beta}$ is 
a double dual Riemann tensor, and 
$F_{\mu \nu}=\partial_{\mu}A_{\nu}-\partial_{\nu}A_{\mu}$ 
is a gauge-field strength. 
We call this interacting Lagrangian ${\cal L}_{\rm HVT}$
the Horndeski-vector-tensor (HVT) term. 
Under a shift $A_{\mu} \to A_{\mu}+\partial_{\mu}\chi$, 
the HVT term respects the $U(1)$ gauge invariance. 
If the $U(1)$ gauge symmetry is broken, it is possible to 
construct a generalized version of massive Proca theories 
with second-order field equations of 
motion (dubbed generalized Proca 
theories) \cite{Heisenberg:2014rta,Tasinato:2014eka,BeltranJimenez:2016rff,Allys:2016jaq}.

In Einstein-Maxwell theory with the $U(1)$ gauge-invariant 
HVT term, Horndeski derived an electrically charged BH 
solution on the static and spherically symmetric background \cite{Horndeski:1978ca}. 
The readers may refer 
to \cite{Mueller-Hoissen:1988cpx,Balakin:2007am,Verbin:2020fzk} for more detailed analyses 
of the background BH solution and also 
to \cite{Esposito-Farese:2009wbc,Barrow:2012ay,BeltranJimenez:2013btb} 
for the cosmological application of the same theory.
The coupling $\beta$ affects the electric field strength 
and gives rise to a nontrivial BH solution 
different from the Reissner-Nordstr\"om (RN) geometry.
In the context of generalized Proca theories where the coupling 
$\beta$ in the HVT term is promoted to a function of 
$X=-A_{\mu}A^{\mu}/2$ with an additional vector-field 
interaction, there are also electrically charged BH 
solutions with a vanishing longitudinal vector 
mode \cite{Heisenberg:2017xda,Heisenberg:2017hwb}. 
We note that, in generalized Proca theories containing 
the HVT Lagrangian as a specific case, the linear 
stability of electrically 
charged BHs was studied for odd-parity perturbations 
in Ref.~\cite{Kase:2018voo}. 
This allows us to rule out some BH solutions 
or to put constraints on the parameter 
space of theories in which the BHs suffer from 
neither ghosts nor Laplacian 
instabilities \cite{Chagoya:2016aar,Minamitsuji:2016ydr,Babichev:2017rti}.
The linear stability conditions of electrically charged 
BHs against even-parity perturbations were not derived 
yet for Einstein-Maxwell-HVT theory or 
generalized Proca theories.

In Einstein-Maxwell theory, the magnetically charged 
BH is also a solution to the gravitational field equations 
of motion. Moreover, the presence of magnetic 
monopoles is ubiquitous in unified gauge  
theory \cite{tHooft:1974kcl,Polyakov:1974ek}
as well as in string theory \cite{Wen:1985qj}.
The lack of observational evidence for magnetic monopoles 
so far may be attributed to the difficulty of pair creation 
due to their heavy masses exceeding the 
order $10^2$~GeV \cite{CDF:2005cvf,Fairbairn:2006gg}.
Still, we cannot exclude the possibility of the existence 
of magnetically charged BHs in Nature. 
Indeed, such BHs may be formed as a result of 
the gravitational clustering of magnetic 
monopoles \cite{Lee:1991vy,Ortiz:1991eu}.
In addition, the possible existence of primordial BHs 
in the early Universe could absorb 
magnetic monopoles \cite{Stojkovic:2004hz,Kobayashi:2021des, Das:2021wei,Estes:2022buj,Zhang:2023tfv}. 
Since the neutralization of magnetic BHs with 
ordinary matter does not occur 
in conductive media, it can be a long-lived stable 
configuration in comparison to the 
electric BH \cite{Maldacena:2020skw, Bai:2020ezy}.

In Einstein-Maxwell theory, the magnetic BH with a charge 
$q_M$ has a duality with the electric BH with a charge $q_E$. 
The background static and spherically symmetric BH solution 
in the former can be simply obtained by changing $q_E$ in the latter
RN solution to $q_M$. At the level of perturbations, 
the quasinormal modes of electrically and 
magnetically charged BHs with the same total charge are equivalent 
to each other. The readers may refer to \cite{Moncrief:1974gw, Moncrief:1974ng, Moncrief:1975sb,Zerilli:1974ai} for 
the perturbation theory of RN BHs and the computation 
of associated quasinormal modes \cite{Gunter:1980,Kokkotas:1988fm,Leaver:1990zz,Berti:2003zu}.
Even for a BH with mixed electric and magnetic 
charges (dyon BHs \cite{Kasuya:1981ef}), 
so long as the BH mass is given, 
the quasinormal modes are solely determined by 
the total BH charge \cite{DeFelice:2023rra}. 
This is attributed to the fact that the linear perturbation 
equations of motion for dyon BHs can decouple into 
two generalized even- and odd-sectors \cite{Pereniguez:2023wxf}. 
The presence of electric-magnetic duality in Einstein-Maxwell theory also leads to the isospectrality of quasinormal modes 
between the two sectors.

There are several ways of breaking the electric-magnetic 
duality in generalized Einstein-Maxwell theories. 
One is to incorporate nonlinear functions of the electromagnetic field strength \cite{Nomura:2020tpc,Nomura:2021efi} like those 
appearing in Born-Infeld theory \cite{Born:1934gh}. 
The other is to introduce electrically charged fields interacting with dyon BHs \cite{Pereniguez:2024fkn}.
The presence of an axion field $\phi$ coupled to 
the electromagnetic field in the form 
$\phi F_{\mu \nu} \tilde{F}^{\mu \nu}$, where 
$\tilde{F}^{\mu \nu}$ is a dual of $F_{\mu \nu}$, 
can realize dyon BHs \cite{Lee:1991jw} 
with quasinormal modes depending on the ratio 
between the magnetic and total 
charges \cite{DeFelice:2024eoj}. 
The presence of the HVT term in Einstein-Maxwell theory 
also breaks the electric-magnetic duality for the 
background charged BH solutions \cite{Horndeski:1978ca,Verbin:2020fzk}.

In this paper, we will study the BH perturbations in 
Einstein-Maxwell theory with the HVT interaction by 
paying particular attention to the breaking of electric-magnetic 
duality for the perturbations of electric and magnetic BHs.
Since the BH perturbations 
in this theory were only studied for electrically 
charged BHs in the odd-parity sector, we will extend 
the analysis to dyon BHs with both odd- and even-parity 
perturbations taken into account. 
For this purpose, in Sec.~\ref{backsec}, we will first 
revisit the background BH solutions in considerable detail and 
highlight the difference between the electric and magnetic BHs.
After deriving the second-order action of perturbations 
for dyon BHs in Sec.~\ref{BHpersec}, we will exploit it 
to derive linear stability conditions (absence of ghosts 
and Laplacian instabilities) for electric 
and magnetic BHs in Secs.~\ref{eleBH} and \ref{magBH}, 
respectively. In both cases, there are four dynamical perturbations arising from the gravitational sectors (two) and 
the vector-field sectors (two). 
In the large-multipole limit ($l \gg 1$), the four angular 
propagation speeds are different from each other 
for both electric and magnetic BHs. 
This suggests the breaking of eikonal correspondence 
between the peak position of potentials of dynamical 
perturbations and the radius of photon sphere \cite{Cardoso:2008bp}. 
Moreover, the no-ghost conditions and the radial/angular 
propagation speeds exhibit differences 
between the electric and magnetic BHs. 
This reflects the breaking of electric-magnetic duality 
for linear perturbations. Finally, we give a summary 
in Sec.~\ref{consec}.

Throughout the paper, we will use the natural unit in which 
the speed of light $c$, the reduced Planck constant $\hbar$, 
and the Boltzmann constant $k_B$ are equivalent to 1.

\section{Background BH solutions in Einstein-Maxwell-HVT theory}
\label{backsec}

We consider theories given by the action 
\be
{\cal S}=\int {\rm d}^4 x \sqrt{-g} \left( \frac{\Mpl^2}{2}R
-\frac{1}{4}F_{\mu \nu}F^{\mu \nu}
+\beta L^{\mu \nu \alpha \beta} F_{\mu \nu}F_{\alpha \beta} 
\right)\,,
\label{action}
\ee
where $g$ is a determinant of the metric tensor 
$g_{\mu \nu}$, $\Mpl$ is the reduced Planck mass, 
$R$ is the Ricci scalar, and $F_{\mu \nu}=
\nabla_{\mu}A_{\nu}-\nabla_{\nu}A_{\mu}$ with 
$A_{\mu}$ being a vector field ($\nabla_{\mu}$ is 
a covariant derivative operator), $\beta$ is a coupling 
constant, and $L^{\mu \nu \alpha \beta}$ is 
the double dual Riemann tensor defined by
\be
L^{\mu\nu\alpha\beta}=\frac{1}{4}
\mathcal{E}^{\mu\nu\rho\sigma}
\mathcal{E}^{\alpha\beta\gamma\delta} R_{\rho\sigma\gamma\delta}\,.
\ee
Here, $R_{\rho\sigma\gamma\delta}$ is the Riemann tensor, 
and $\mathcal{E}^{\mu\nu\alpha\beta}$ is the 
anti-symmetric Levi-Civita tensor\footnote{Alternatively, 
we may introduce $\epsilon^{\mu\nu\rho\sigma}$ and 
$\epsilon_{\mu\nu\rho\sigma}$ as
$\mathcal{E}^{\mu\nu\rho\sigma}=\epsilon^{\mu\nu\rho\sigma}
/\sqrt{-g}$ and 
$\mathcal{E}_{\mu\nu\rho\sigma}=\sqrt{-g}\,
\epsilon_{\mu\nu\rho\sigma}$, so that 
$\epsilon^{0123}=-1$ and $\epsilon_{0123}=1$. 
Then, we can write the HVT term ${\cal L}_{\rm HVT}=\beta L^{\mu \nu \alpha \beta} F_{\mu \nu}F_{\alpha \beta}$ in the form 
${\cal L}_{\rm HVT}=(\beta/4) \epsilon^{\mu\nu\rho\sigma}
\epsilon_{\alpha\beta\gamma\delta} R_{\rho\sigma}{}^{\gamma\delta}F_{\mu\nu} F^{\alpha\beta}$.} with the components 
$\mathcal{E}^{0123}=-1/\sqrt{-g}$ 
and $\mathcal{E}_{0123}=\sqrt{-g}$.
In Einstein-Maxwell-HVT theory given by the action (\ref{action}), which respects the $U(1)$ gauge invariance, 
the field equations of motion 
are kept up to second order \cite{Horndeski:1978ca}.

Varying the action \eqref{action} with respect to $A_\mu$, one obtains the modified Maxwell equations
\be
\nabla_\mu\left(F^{\mu\nu}-4\beta F_{\alpha\beta}L^{\alpha\beta\mu\nu}\right)=0\,.
\label{modifiedMW}
\ee
On the other hand, variation with respect to 
$g_{\mu \nu}$ gives the gravitational field equations
\be
\Mpl^2 G_{\mu\nu}-{F_\mu}^\beta F_{\nu\beta}+\frac{1}{4}g_{\mu\nu}F_{\alpha\beta}F^{\alpha\beta}+\beta H_{\mu\nu}=0\,,
\label{graveq}
\ee
where
\be
H_{\mu\nu}\equiv g_{\mu\nu}L^{\alpha\beta\gamma\delta}F_{\alpha\beta}F_{\gamma\delta}+2\tilde{F}_{\mu\sigma}\tilde{F}^{\gamma\delta}{R^\sigma}_{\nu\gamma\delta}-4\nabla^\sigma{\tilde{F}_{\gamma\nu}}\nabla^\gamma\tilde{F}_{\mu\sigma}-4\tilde{F}_{\gamma\nu}\left(R^{\rho\gamma}\tilde{F}_{\mu\rho}+{R^\gamma}_{\sigma\rho\mu}\tilde{F}^{\rho\sigma}\right)\,.
\ee
Here, $G_{\mu\nu}$ is the Einstein tensor 
and $\tilde{F}^{\mu\nu}\equiv F_{\alpha\beta}\mathcal{E}^{\alpha\beta\mu\nu}/2$ 
is the dual strength tensor.

Let us consider a static and spherically symmetric 
background given by the line element 
\be
\rd s^2=-f(r) \rd t^{2} +h^{-1}(r) \rd r^{2}+ 
r^{2} \left( \rd \theta^{2}+\sin^{2}\theta\,\rd\varphi^{2} 
\right)\,,
\label{metric_bg}
\ee
where $f$ and $h$ are functions of the radial coordinate $r$. 
For the vector field, we choose the following configuration
\be
A_{\mu}=\left[ A_0 (r), 0, 0, -q_M \cos \theta \right]\,,
\ee
where $A_0$ is a function of $r$, and $q_M$ is a constant 
corresponding to the magnetic charge. 
Due to the presence of the $U(1)$ gauge symmetry, 
we set the radial vector component $A_1(r)$ zero.

Computing the action (\ref{action}) for the metric ansatz (\ref{metric_bg}) and varying it with respect to $f$, $h$, and $A_0$, respectively, we obtain the following field 
equations of motion 
\ba
& &
\Mpl^2 r h'+\Mpl^2 \left( h-1 \right)+\frac{q_M^2}{2r^2}
+\frac{r^2 h}{2f}A_0'^2+\frac{4\beta}{r^4 f} \left[ 
q_M^2 (rh'-6h)f-r^4 h(h-1)A_0'^2 \right]=0\,,
\label{back1}\\
& &
\Mpl^2 r f'+\frac{\Mpl^2 f(h-1)}{h}+\frac{q_M^2f}{2r^2h}
+\frac{r^2}{2}A_0'^2+\frac{4\beta}{r^3} \left( 
q_M^2 f'+r^3A_0'^2-3r^3h A_0'^2 \right)=0\,,
\label{back2}\\
& & 
\left( \frac{\sqrt{h}[r^2-8 \beta (h-1)]}{\sqrt{f}}A_0' \right)'=0\,,
\label{back3}
\ea
where a prime represents the derivative with respect to $r$.
We can integrate Eq.~(\ref{back3}) to give 
\be
A_0'(r)=\frac{\sqrt{f}\,q_E}{\sqrt{h}[r^2-8 \beta (h-1)]}\,,
\label{back4}
\ee
where $q_E$ is a constant corresponding to the electric charge.
Substituting Eq.~(\ref{back4}) into Eqs.~(\ref{back1}) and (\ref{back2}), 
we obtain the first-order differential equations for $h$ and $f$.
In the limit that $\beta \to 0$, the solutions respecting the boundary conditions $f \to 1$ and $h \to 1$ at spatial 
infinity correspond to the RN BHs with 
$f=h=1-2M/r+(q_E^2+q_M^2)/(2\Mpl^2 r^2)$, where $M$ is the 
Arnowitt-Deser-Misner (ADM) mass. In this case, the metric components are not modified by the simultaneous changes 
$q_E \to q_M$ and $q_M \to q_E$ due to a duality between the electric and magnetic charges. 
For the purely electrically charged BH ($q_E \neq 0$ and $q_M=0$), the nonvanishing temporal vector component, which contains 
the $\beta$ dependence, affects the metrics through the $A_0'^2$ terms in Eqs.~(\ref{back1}) and (\ref{back2}).
For the purely magnetically charged BH ($q_M \neq 0$ and $q_E=0$), 
we have $A_0'=0$, but the gravitational Eqs.~(\ref{back1}) and (\ref{back2}) contain the dependences of $q_M$ and $\beta$.

For $\beta \neq 0$, the electric-magnetic duality 
is already broken at the background level. 
To see this property, let us first discuss the 
purely electrically charged BH. 
In this case, Eqs.~(\ref{back1}) and (\ref{back2}) 
reduce, respectively, to 
\ba
& &
\Mpl^2 r h'+\Mpl^2 \left( h-1 \right)
+\frac{q_E^2}{2[r^2-8 \beta (h-1)]}=0\,,\\
& &
\Mpl^2 r f'+\frac{\Mpl^2 f(h-1)}{h}
+\frac{q_E^2f [r^2-8 \beta (3h-1)]}
{2h [r^2-8 \beta (h-1)]^2}=0\,.
\ea

Let us consider a regime in which the coupling 
constant $\beta$ is close to 0. 
Then, up to linear order in $\beta$, we obtain 
the following integrated solutions 
\ba
& &
h \simeq 1-\frac{2M}{r}+\frac{q_E^2}{2\Mpl^2 r^2}
-\frac{2 q_E^2 (5M \Mpl^2r -q_E^2)}{5\Mpl^4 r^6}\beta\,,
\label{hE}\\
& &
f \simeq 1-\frac{2M}{r}+\frac{q_E^2}{2\Mpl^2 r^2}
+\frac{q_E^2 (10 M \Mpl^2 r - 10 \Mpl^2 r^2 - 3q_E^2)}
{5\Mpl^4 r^6}\beta\,, 
\label{fE}
\ea
which show that the coupling $\beta$ gives rise to 
the difference between $h$ and $f$.
For the purely magnetically charged BH, 
Eqs.~(\ref{back1}) and (\ref{back2}) yield 
\ba
& &
\Mpl^2 r h'+\Mpl^2 \left( h-1 \right)
+\frac{q_M^2 [r^2+8\beta (rh'-6h)]}{2r^4}=0\,,\\
& &
\Mpl^2 r f'+\frac{\Mpl^2 f(h-1)}{h}
+\frac{q_M^2(rf+8 \beta f' h)}{2r^3 h}=0\,,
\ea
respectively. 
On using the expansion of $\beta$ around 0 
in the small-coupling regime, the solutions to $h$ 
and $f$, up to the linear order in $\beta$, are given by 
\ba
& &
h \simeq 1-\frac{2M}{r}+\frac{q_M^2}{2\Mpl^2 r^2}
+\frac{2q_M^2 (35 M \Mpl^2 r -20 \Mpl^2 r^2 
-8q_M^2)}{5\Mpl^4 r^6}\beta\,,
\label{hM}\\
& &
f \simeq 1-\frac{2M}{r}+\frac{q_M^2}{2\Mpl^2 r^2}
+\frac{q_M^2 (10 M \Mpl^2 r -10 \Mpl^2 r^2 
-q_M^2)}{5\Mpl^4 r^6}\beta\,.
\label{fM}
\ea
Thus, the difference between $h$ and $f$ is induced by 
the coupling $\beta$. 
Applying the change $q_E \to q_M$ in Eqs.~(\ref{hE}) and (\ref{fE}), the resulting metric components are not equivalent to Eqs.~(\ref{hM}) and (\ref{fM}), respectively.
This means that, for the small coupling $\beta$, 
the electric-magnetic duality does not hold 
even at the background level.

\subsection{Expansion around the horizon}

In the following, we will study the properties of background 
dyon BH solutions without using an approximation of the small coupling $\beta$. For this purpose, we first consider the BHs 
with at least one of the event horizons. 
In Sec.~\ref{parasec}, we will also study the parameter space 
in which the horizons disappear.
In the vicinity of the outer horizon characterized by 
the radius $r_h$, we expand $f$ and $h$ in the forms 
\be
f(x)=\sum_{i=1} f_i (x-1)^i\,,\qquad 
h(x)=\sum_{i=1} h_i (x-1)^i\,,
\label{fxho}
\ee
where $x \equiv r/r_h$, and $f_i$, $h_i$ are constants. 
Upon using Eq.~(\ref{back4}), we can eliminate the 
$A_0'$-dependent terms in Eqs.~(\ref{back1}) and 
(\ref{back2}). Then, the coefficients $f_i$ and $h_i$ 
are known order by order. 
While $f_1$ is undetermined, $h_1$ is fixed to be 
\be
h_1=\frac{2-\tilde{q}_E^2-\tilde{q}_M^2
+8\tilde{\beta} (2-\tilde{q}_M^2)}
{2(1+8\tilde{\beta})(1+4 \tilde{\beta} \tilde{q}_M^2)}\,,
\ee
where 
\be
\tilde{\beta} \equiv \frac{\beta}{r_h^2}\,,\qquad 
\tilde{q}_E \equiv \frac{q_E}{\Mpl r_h}\,,\qquad 
\tilde{q}_M \equiv \frac{q_M}{\Mpl r_h}\,.
\ee
Equivalently, one can replace the previous 
definition of 
$r_h$ with a general value $r_{\rm pivot}$, and then one fixes $r_h$ (in terms of $r_{\rm pivot}$) to have the ADM mass $M$ equal to unity. For the electric and magnetic BHs,
we have that 
$h_1=(2-\tilde{q}_E^2+16 \tilde{\beta})/
[2(1+8\tilde{\beta})]$ and 
$h_1=(2-\tilde{q}_M^2)
/[2(1+4 \tilde{\beta} \tilde{q}_M^2)]$, 
respectively. Changing $\tilde{q}_E$ to $\tilde{q}_M$ for 
the former, the resulting value of $h_1$ is not equivalent 
to the latter. Thus, the electric-magnetic duality 
does not hold for $h_1$.
Other coefficients like $f_2$ and $h_2$ are also 
known accordingly. 
Since $h(x)>0$ outside the horizon ($x>1$), 
we require that $h_1>0$, which translates to 
the condition
\be
(1+8\tilde{\beta})(1+4 \tilde{\beta} \tilde{q}_M^2) 
\left[ 2-\tilde{q}_E^2-\tilde{q}_M^2
+8\tilde{\beta} (2-\tilde{q}_M^2) \right]>0\,.
\label{qcon}
\ee
For $\tilde{\beta}>0$, the inequality (\ref{qcon}) 
is satisfied if $2-\tilde{q}_E^2-\tilde{q}_M^2
+8\tilde{\beta} (2-\tilde{q}_M^2)>0$. 
This condition translates to 
$\tilde{q}_E^2<2(1+8 \tilde{\beta})$ 
for the electric BH and 
$\tilde{q}_M^2<2$ 
for the magnetic BH. 
For the numerical purpose, we expand $f$ and $h$ up 
to the $i=2$ order in Eq.~(\ref{fxho}) and then use 
them as the boundary 
conditions in the vicinity of the outer horizon. 

\subsection{Expansion at spatial infinity}

At spatial infinity, we impose the boundary conditions 
$f \to 1$ and $h \to 1$. In this regime, the radial 
derivative of $A_0$, which corresponds to the electric 
field, decreases as $A_0' \to q_E/r^2$ 
for $q_E \neq 0$.
Far outside the outer horizon, we also expand $f$ and $h$ 
in the forms $f=1+\sum_{i=1} F_i r^{-i}$ and 
$h=1+\sum_{i=1} H_i r^{-i}$, where $F_i$ and $H_i$ 
are constants. Up to the order of $r^{-6}$, we obtain 
the following large-distance solutions 
\ba
f &=& 1-\frac{2M}{r}+\frac{q_E^2+q_M^2}{2 \Mpl^2 r^2}
-\frac{2\beta(q_E^2+q_M^2)}{\Mpl^2 r^4}
+\frac{2\beta M(q_E^2+q_M^2)}{\Mpl^2 r^5}
-\frac{\beta (q_E^2+q_M^2) (3q_E^2+q_M^2)}{5\Mpl^4 r^6}
+{\cal O} (r^{-7})\,,\label{finf}\\
h &=& 1-\frac{2M}{r}+\frac{q_E^2+q_M^2}{2 \Mpl^2 r^2}
-\frac{8\beta q_M^2}{\Mpl^2 r^4}
-\frac{2\beta M(q_E^2-7q_M^2)}{\Mpl^2 r^5}
+\frac{2\beta (q_E^2+q_M^2) (q_E^2-8q_M^2)}{5\Mpl^4 r^6}
+{\cal O} (r^{-7})\,,\label{hinf}
\ea
where $M$ is a constant corresponding to the BH ADM mass. In the absence of the magnetic charge, the expanded solutions (\ref{finf}) and (\ref{hinf}) coincide with those derived in Ref.~\cite{Heisenberg:2017hwb}. Up to the order of $r^{-2}$, the large-distance metric components are equivalent to those of the RN metric. For $q_M \neq 0$, the corrections from the coupling $\beta$ to $f$ and $h$ start to appear at the order of $r^{-4}$.
If $q_M=0$, the term of order $r^{-4}$ in $h$ vanishes, and hence the difference between the purely electrically and magnetically charged BHs appears at this order. For the metric component $f$, the electric-magnetic duality is broken at the order of $r^{-6}$.

\subsection{Numerical BH solutions with event horizons}

To study whether the solutions to $f$ and $h$ 
expanded around the horizon connect to those at 
spatial infinity, we numerically integrate 
Eqs.~(\ref{back1}) and (\ref{back2}) with Eq.~(\ref{back4}) 
toward larger distances by using Eq.~(\ref{fxho}) 
as the boundary conditions. 
For the metric component $f_1$, we choose $f_1=h_1$ for 
the first integration and then obtain the value of $f=f_{\infty}$ 
at a sufficiently large distance. 
Then, in the second integration, we divide $f_1$ by $f_{\infty}$ 
and then solve the differential equations again. 
This leads to the large-distance solutions respecting the 
asymptotic flatness ($f \to 1$, $h \to 1$ as $r \to \infty$).

\begin{figure}[ht]
\begin{center}
\includegraphics[height=3.5in,width=3.4in]{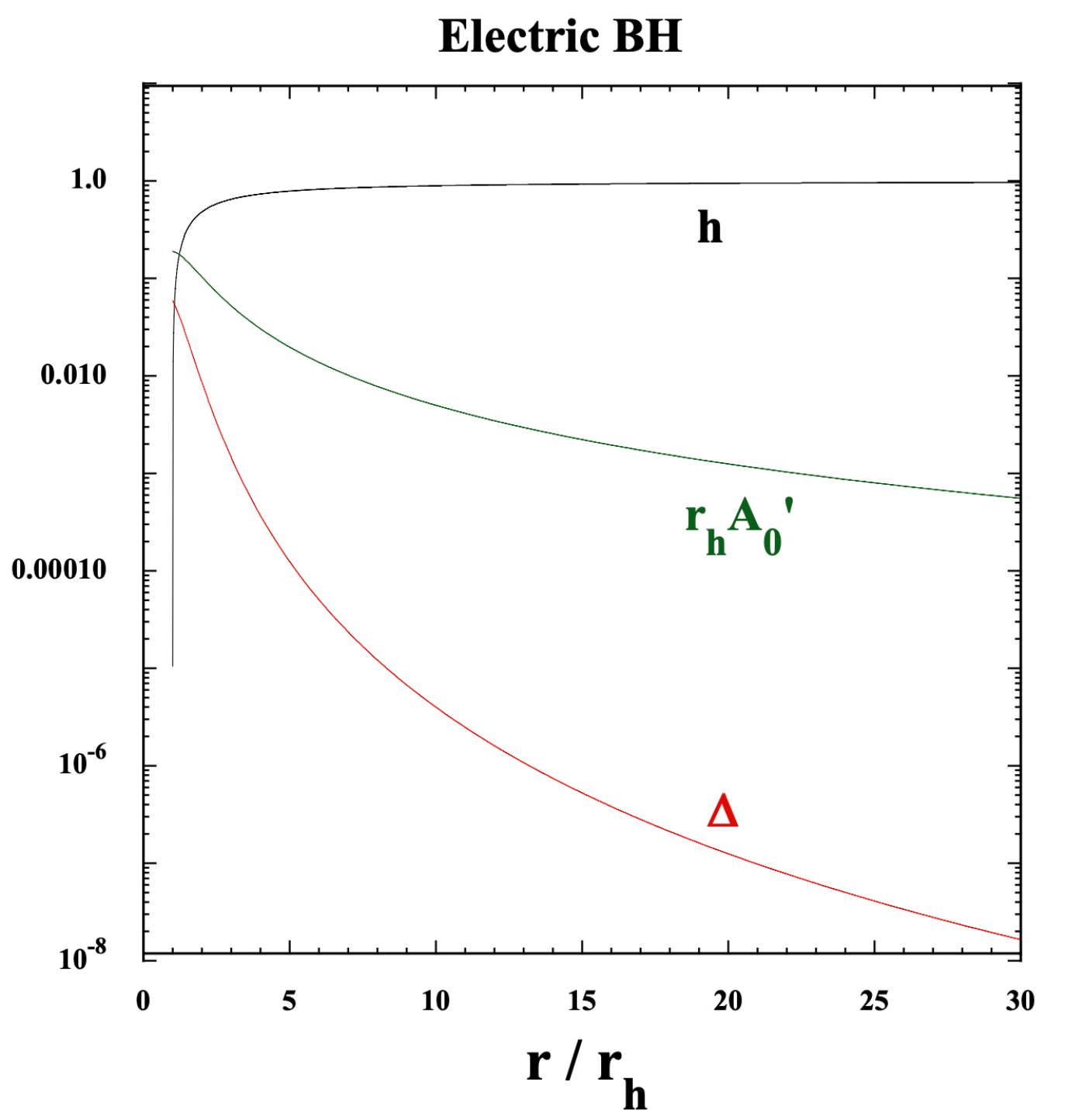}
\includegraphics[height=3.5in,width=3.4in]{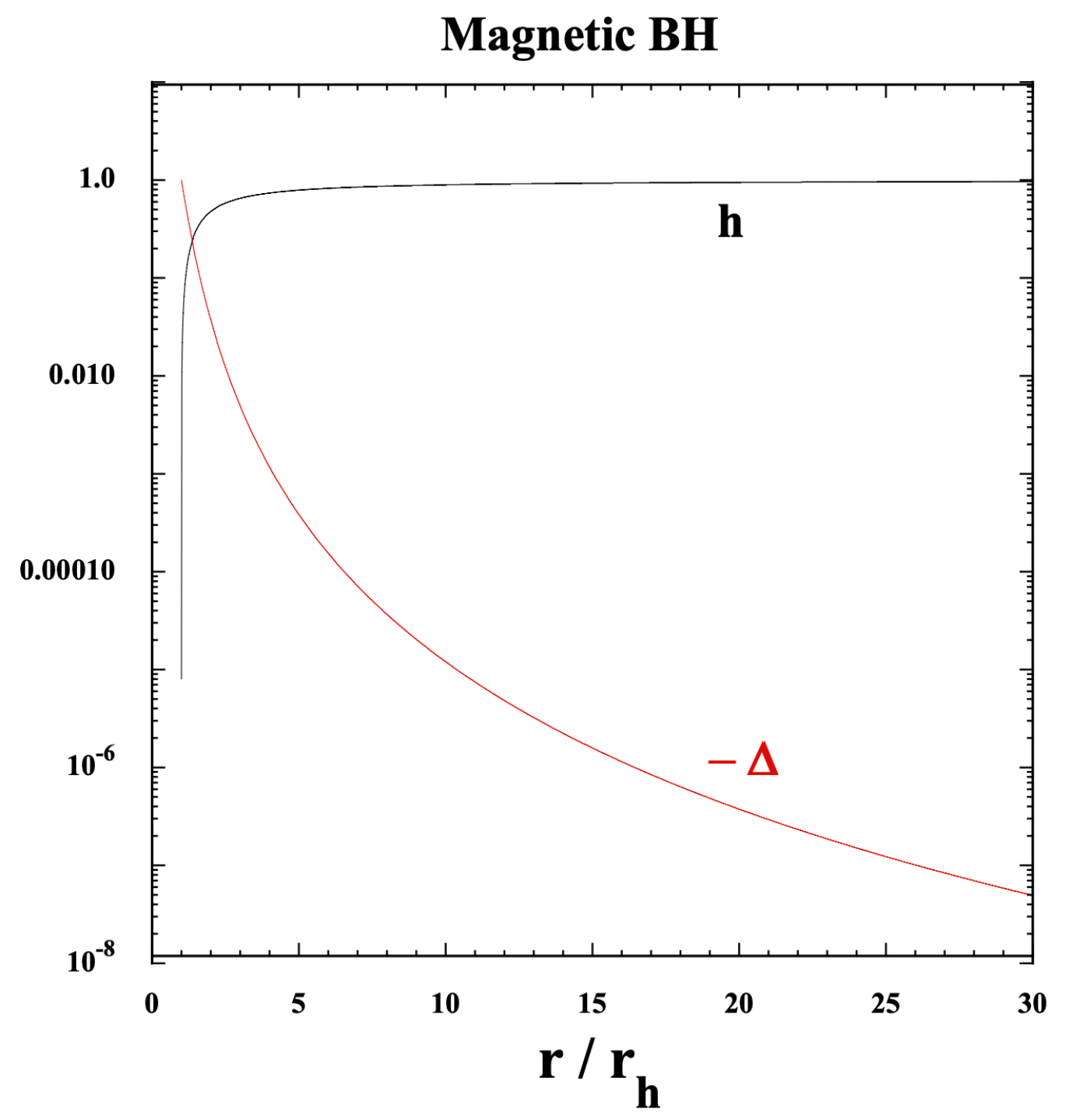}
\end{center}
\caption{\label{fig1} 
The left panel shows $h$, $\Delta=r_h(f'/f-h'/h)$, 
and $r_h A_0'$ versus $r/r_h$ for the electric BH 
with $\tilde{q}_E=0.5$ and $\tilde{\beta}=0.2$. 
The right panel shows $h$ and $-\Delta$ versus $r/r_h$ 
for the magnetic BH with $\tilde{q}_M=0.5$ and $\tilde{\beta}=0.2$. For the magnetic BH, 
$A_0'$ is vanishing.
In both cases, we integrate the background equations 
by using the boundary conditions near the outer horizon (at a distance $r/r_h=1+10^{-6}$). 
For the case $q_M=0$, the ADM mass is 
$2M \simeq 1.0969$, whereas for $q_E=0$, we have $2M \simeq 1.0792$. 
For the numerical computation, the Runge-Kutta method is used 
in Fortran with the step size $10^{-5}$. 
The choice of a smaller step size does not change the results shown 
in this figure.}
\end{figure}

In the left panel of Fig.~\ref{fig1}, we plot 
$h$, $\Delta=r_h(f'/f-h'/h)$, and $r_h A_0'$ 
for the purely electrically charged BH with 
$\tilde{q}_E=0.5$ and $\tilde{\beta}=0.2$. 
The solution around the outer horizon 
joins the large-distance solutions (\ref{finf}) 
and (\ref{hinf}). We note that $\Delta$ 
is nonvanishing for $\beta \neq 0$ and that 
$\Delta$ is largest at $r=r_h$. 
Thus, the nonzero coupling $\beta$ gives rise to 
the difference between $f$ and $h$ especially in the vicinity 
of the horizon. The radial derivative of $A_0$ also 
receives corrections from the coupling $\beta$ 
in the region close to $r=r_h$.
At large distances, it has the asymptotic behavior 
$A_0' \propto r^{-2}$.

In the right panel of Fig.~\ref{fig1}, we show 
$h$ and $-\Delta$ for the purely magnetically 
charged BH with $\tilde{q}_M=0.5$ and $\tilde{\beta}=0.2$. 
Again, the solutions around $r=r_h$ and at large distances 
join each other.
In this case, the quantity $-\Delta$ reaches the order 1 
on the horizon and hence the difference between the two 
metric components is larger in comparison to the electric BH 
with the same charge. We note that $A_0'$ is vanishing for 
the magnetic BH. 

For the same charge and coupling $\beta$, we numerically confirm that the metric components of the electric BH are different from those of the magnetic BH. 
As $|\beta|$ increases, this difference tends to be more significant.
Thus, the HVT coupling breaks the electric-magnetic duality of the background BH solutions outside the outer horizon. 
We also carried out numerical simulations for dyon BHs with mixed electric and magnetic charges and confirmed that, for $\tilde{\beta}$, $\tilde{q}_E$, $\tilde{q}_M$ in the range (\ref{qcon}), there exist regular numerical solutions of $f$ and $h$ connecting the two solutions expanded around the horizon and spatial infinity.

\subsection{Parameter spaces of BHs with/without naked 
singularities}
\label{parasec}

\begin{figure}[ht]
\begin{center}
\includegraphics[height=2.2in,width=3.5in]{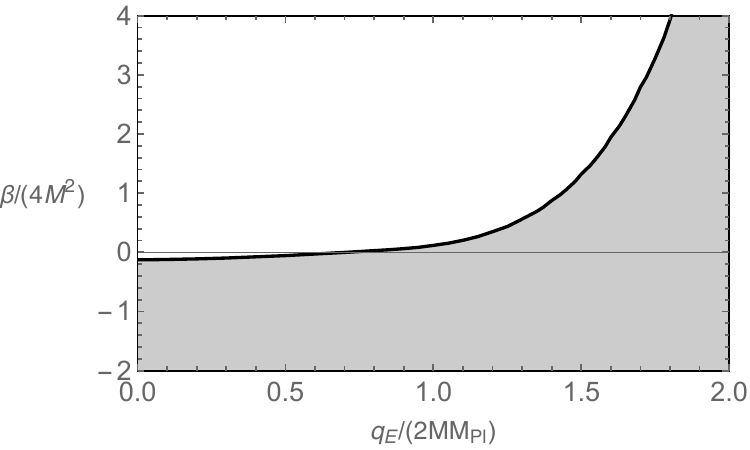}
\includegraphics[height=2.2in,width=3.5in]{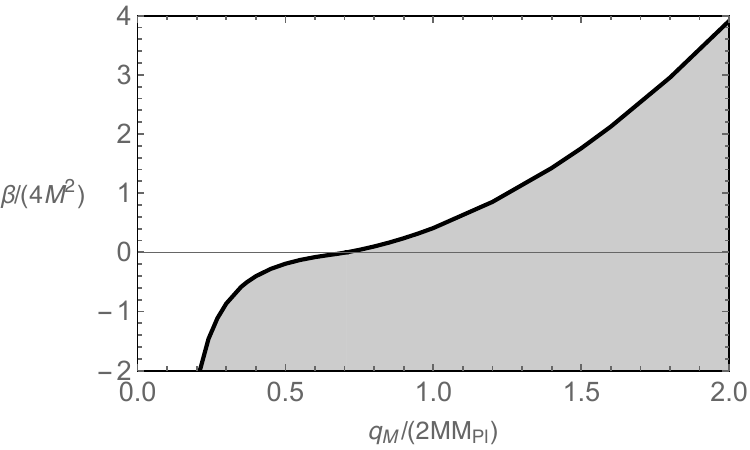}
\end{center}
\caption{\label{fig:parameterspace} 
For the purely electrically charged (left) and magnetically charged (right) BHs, the shaded regions correspond to the parameter spaces in which the horizons disappear with the appearance of naked singularities. To generate this figure, we adopt the step size $10^{-5}$ in the numerical integration as that used in Fig.~\ref{fig1}. Proper integration methods are adaptively chosen by the \textit{NDSolve} command of \textit{Mathematica} to accommodate the stiff systems near the singularity. 
}
\end{figure}

So far, the background BH solution has been investigated by assuming the existence of event horizons. 
Indeed, there exist some ranges of the parameter space in which the event horizons disappear with the appearance of naked singularities. The violation of electric-magnetic duality can also be demonstrated by considering such parameter spaces for the electric and magnetic BHs.
To show this, we take Eqs.~\eqref{finf} and \eqref{hinf} as the boundary conditions at a sufficiently large distance and then numerically integrate Eqs.~(\ref{back1}) and (\ref{back2}) with Eq.~(\ref{back4}) inward. After that, we identify the region of parameter space in which the horizon disappears. We numerically obtain the BH ADM mass according to the formula
\be
M=\lim_{r \to \infty} r(1-h)/2\,.
\ee
In Fig.~\ref{fig:parameterspace}, we show the parameter spaces 
in which a naked singularity appears (shaded region) 
for the electric BH (left) and the magnetic BH (right). In this figure, all the parameters are scaled to $2M$ rather than $r_h$ to account for the parameter spaces in which the solutions do not have event horizons. 
When $\beta=0$, the solid curve, which represents the boundary that distinguishes between the scenarios of BHs and naked singularities, intersects at 
$q_E/(2M M_{\rm pl})=q_M/(2M M_{\rm pl})=1/\sqrt{2}$. This corresponds to the extremal RN solution.

When $\beta>0$, the BH solutions with charges larger than those of the usual extremal RN limit can exist both for the electric and magnetic BHs. 
Inside the horizon, the solutions extend down to $r=0$ for magnetic BHs and for a large portion of the parameter space of electric BHs. In the latter electric BH case, the two metric functions around 
the origin can be approximated as
\be
f(r) \simeq \frac{\tilde{f}_0}{r}\,,\qquad 
h(r) \simeq \frac{\tilde{h}_0}{r}\,.\label{fhorigin}
\ee
The coefficients $\tilde{f}_0$ and $\tilde{h}_0$ are determined by the boundary conditions. 
For BHs, they are both negative, and only one horizon is present. Also, there is a tiny region of the parameter space for electric BHs characterized by $0\le\beta/4M^2<0.117(q_E/2M\Mpl)^6$, in which two horizons exist and the solutions terminate at a singular surface $r>0$ inside the inner one. 
For the magnetic BH, the expanded solutions around $r=0$ are
\begin{align}
f(r)&\simeq \tilde{F}_0\left(\frac{1}{r^4}-\frac{4\Mpl^2}{q_M^2r^2}-\frac{3\Mpl^2\ln{r}}{4q_M^2\beta}+\cdots\right)\,,\nonumber\\ 
h(r) &\simeq \frac{r^2}{32\beta}-\frac{\Mpl^2r^4}{8q_M^2\beta}-\frac{3\Mpl^2 r^6\ln{r}}{128q_M^2\beta^2}+\cdots\,,
\end{align}
where the coefficient $\tilde{F}_0=\sqrt{\beta}|q_M|^3/(4\Mpl^3)$ is determined by requiring that $f(\infty)/h(\infty)=1$. 
Although $h$ is finite in the limit $r \to 0$, the Ricci scalar diverges as $R \simeq 2/r^2$.
The coefficient $\tilde{F}_0$ for the magnetic BH is positive and there is an inner horizon inside the BH. 

On the other hand, when $\beta<0$, the allowed range of $q_M$ for the magnetic BH shrinks. Furthermore, the allowed region of negative $\beta$ for the electric BH is substantially limited. 
Also, in the magnetic case, the solutions extend down to a singular surface with a radius $r>0$ (see Sec.~\ref{sec:exactqm} for more details on the magnetic BH), while the solutions in the electric case extend down to the origin. In this case, the metric functions near the origin can be approximated as Eqs.~\eqref{fhorigin}, with positive coefficients $\tilde{f}_0$ and $\tilde{h}_0$. 
Hence, there are two horizons. 
Note that, for the electric BH (left), the solid curve converges to $|\beta|\rightarrow M^2/2$ when $|q_E|\rightarrow 0^+$.

%
\subsection{Exact background solutions for 
magnetic BHs}\label{sec:exactqm}

In the purely magnetically charged case ($q_M \neq 0$ and $q_E=0$), it is possible to obtain the background BH solutions analytically.  
For this purpose, we define two radii $r_\beta$ and $r_M$ as
\be
\beta =\epsilon_{\beta}\frac{\Mpl^{2}r_{\beta}^{4}}{4q_{M}^{2}}\,,\qquad
|q_{M}| =\sqrt{2}r_{M}\Mpl\,,
\ee
where $\epsilon_{\beta}=+1$ for $\beta>0$ and $\epsilon_{\beta}=-1$ for $\beta<0$. 
Then, there is the following relation 
\be
r_{\beta}=\left( \frac{4|\beta|q_{M}^{2}}{\Mpl^{2}}\right)^{1/4}=
\left( 8|\beta| \right)^{1/4}\sqrt{r_{M}}\,.
\ee
The solution to Eq.~(\ref{back1}) with $q_E=0$ and $A_0'=0$ 
can be expressed as 
\be
h=\frac{r^{6}}{(r^{4}+\epsilon_{\beta}r_{\beta}^{4})^{7/4}}\int_{r_{h}}^{r}\frac{(r_{1}^{2}-r_{M}^{2})(r_{1}^{4}+\epsilon_{\beta}r_{\beta}^{4})^{3/4}}{r_{1}^{5}}{\rm d}r_{1}\,,
\label{hana}
\ee
where $r_h$ corresponds to the position of an apparent horizon seen by an observer at infinity.
The metric component $h$ satisfies
\be
h(r_{h})=0\,,\qquad {\rm and} 
\qquad h(r>r_h)>0\,.
\ee
We note that the integral on the right-hand side of Eq.~(\ref{hana}) can be further expressed in terms of hypergeometric functions \cite{Verbin:2020fzk}. 
For the validity of the solution (\ref{hana}), we are assuming that the horizon radius $r_h$ is in the range $r_h\geq r_M$ (if $\beta \geq 0$), or $r_{h}>{\rm max}(r_{M},r_{\beta})$ (if $\beta<0$). 
For $r_h$ of the same order as $r_s=2G_N \tilde{M}$, where $G_N=1/(8 \pi \Mpl^2)$ is the gravitational constant and $\tilde{M}$ is the ADM mass of the system (with the actual 
mass dimension), the inequality $r_s \approx r_h \geq r_M$ translates to $\tilde{M}/\Mpl \gtrsim 4\pi |q_M|/\sqrt{2}$. 
At least for astrophysical BHs, the inequalities $\tilde{M}/\Mpl\gg |q_M|$ or $r_h \gg r_M$ is trivially verified. 
For $\beta<0$, the other inequality $r_h>r_{\beta}$ translates to 
\be
\frac{(-8 \beta)^{1/4}}{\sqrt{r_h}}
<\sqrt{\frac{r_h}{r_M}}\,,
\label{betaeq}
\ee
which gives the upper limit on $-\beta$. 
So long as $r_h \gg r_M$, the inequality (\ref{betaeq}) is satisfied for $|\beta|/r_h^2 \lesssim 1$.

We can find the solution for $f$ as
\be
f=C_{f}\exp\!\left[-\int^{r}_{r_s} 
\frac{r[r^{2}(h-1)+r_{M}^{2}]}{(r^{4}+\epsilon_{\beta}r_{\beta}^{4})\,h}\,{\rm d}r\right]\,,
\ee
where $C_{f}$ is an integration constant.
When $r\to r_{h}^{+}$, then we have $h\to0^{+}$ and also $f\to0$, in general. 

For $r\gg r_{h}$, the metric component (\ref{hana}) has the asymptotic behavior
\be
h\approx\frac{1}{r} \int^{r}_{r_s} 
{\rm d}r_{1}
=1-\frac{r_s}{r}\,.
\ee
In this same limit, we have
\be
f\approx C_{f}\exp\left( \int^{r}_{r_s} \frac{r_{s}}{r^{2}}\,{\rm d}r \right)={\tilde C}_{f}\exp\!
\left(-\frac{r_{s}}{r}\right) \approx {\tilde C}_{f}\left(1-\frac{r_{s}}{r}\right)\,,
\ee
where the integration constant 
${\tilde C}_{f}=C_f\,e$ can be set to 1 to obtain the solution $f\approx1-r_{s}/r$.

For $\epsilon_{\beta}=-1$, i.e., $\beta<0$, we consider the limit $r \to r_{\beta}^{+}$ in $h$.
Setting $r=r_{\beta}\,(1+\epsilon)$ with $\epsilon \to 0^{+}$ in Eq.~(\ref{hana}), it follows that 
\be
|h|\approx\frac{\sqrt{2}}{16r_{\beta}\epsilon^{7/4}}\left\vert\int_{r_{h}}^{r_{\beta}}\frac{(r_{1}^{2}-r_{M}^{2})(r_{1}^{4}-r_{\beta}^{4})^{3/4}}{r_{1}^{5}}{\rm d}r_{1}\right\vert\to\infty\,.
\ee
In the same limit, the Ricci scalar behaves as
\be
\lim_{r\to r_\beta^{+}}|R|=\lim_{r\to r_\beta^{+}}\frac{2r_{\beta}^{4}\,|3h\,r^{4}-7hr_{\beta}^{4}-r^{2}r_{M}^{2}+r_{\beta}^{4}|}{r^{2}\,(r^{4}-r_{\beta}^{4})^{2}}=\infty\,.
\ee
Therefore, for $\beta<0$, we reach a singularity located at $r_\beta>0$. 
This does not correspond to a naked singularity when $r_h\gg r_M$ and $|\beta|/r_h^2\lesssim1$, see the parameter space in the right panel of Fig.~\ref{fig:parameterspace}.

\section{Black hole perturbations}
\label{BHpersec}

We proceed to the discussion of BH perturbations on the static and spherically symmetric background given by the line element (\ref{metric_bg}). 
Since we are interested in the linear stability of BHs outside the outer horizon, we will assume that $f(r)>0$ and $h(r)>0$. Without loss of generality, we consider the $m=0$ components of spherical harmonics $Y_{lm}(\theta, \varphi)$, i.e., $Y_l(\theta) \equiv Y_{l0}(\theta)$. Metric perturbations $h_{\mu \nu}$ have the following components \cite{Regge:1957td,Zerilli:1970se,Zerilli:1970wzz}
\ba
\hspace{-0.7cm}
& &
h_{tt}=f(r) H_0(t,r) Y_{l}(\theta)\,,
\qquad
h_{tr}=h_{rt}=H_1(t,r) Y_{l}(\theta)\,,
\qquad
h_{t \theta}=h_{\theta t}=h_0(t,r)Y_{l, \theta} (\theta)\,,
\nonumber \\
\hspace{-0.7cm}
& &
h_{t \varphi}=h_{\varphi t}
=-Q(t,r) (\sin \theta) Y_{l, \theta} (\theta), 
\qquad
h_{rr}=h^{-1}(r) H_2(t,r) Y_{l}(\theta)\,,
\qquad
h_{r \theta}=h_{\theta r}
=h_1 (t,r)Y_{l, \theta}(\theta),\nonumber \\
\hspace{-0.7cm}
& &
h_{r \varphi}=h_{\varphi r}=-W(t,r) (\sin \theta) Y_{l,\theta} (\theta)\,,\qquad
h_{\theta \theta}=r^2 K(t,r)Y_{l}(\theta)
+r^2 G(t,r)Y_{l,\theta \theta}(\theta)\nonumber \\
\hspace{-0.7cm}
& &
h_{\varphi \varphi}=r^2 K(t,r) (\sin^2 \theta) 
Y_{l}(\theta)+r^2G(t,r) (\sin \theta)(\cos \theta) Y_{l,\theta}(\theta)\,,\qquad
h_{\theta \varphi}=\frac{1}{2}U(t,r) 
[(\cos \theta) Y_{l,\theta} (\theta)
-(\sin \theta) Y_{l,\theta \theta} (\theta)],
\label{hcom}
\ea
where the summation of $Y_{l} (\theta)$ with respect to the multipoles $l$ is omitted, and we use the notations $Y_{l,\theta}=\rd Y_l/\rd \theta$ and $Y_{l,\theta \theta}=\rd^2 Y_l/\rd \theta^2$. 
The vector field $A_{\mu}$ has the following perturbed components \cite{Kase:2023kvq}
\be
\delta A_t=\delta A_0 (t,r) Y_{l}(\theta),\qquad 
\delta A_r=\delta A_1 (t,r) Y_{l}(\theta),\qquad
\delta A_\theta=0,\qquad 
\delta A_{\varphi}=-\delta A(t,r) (\sin \theta) 
Y_{l,\theta}(\theta)\,,
\label{perma}
\ee
where we have chosen the gauge $\delta A_\theta=0$ due to the existence of a $U(1)$ gauge symmetry.
The four fields $Q$, $W$, $U$, $\delta A$ correspond to the perturbations in the odd-parity sector, whereas the nine fields $H_0$, $H_1$, $H_2$, $h_0$, $h_1$, $K$, $G$, $\delta A_0$, $\delta A_1$ are the perturbations in the even-parity sector.
For $l=0$ and $l=1$, we have $Y_{0}=1/\sqrt{4\pi}$ and $Y_{1}=\sqrt{3/(4\pi)} \cos \theta$, respectively, in which cases $h_{\theta \varphi}=0$. 
Then, the contributions from the perturbation $U(t,r)$ to the action appear only for the multiples $l \geq 2$.

Let us consider an infinitesimal gauge transformation $x_{\mu} \to x_{\mu}+\xi_{\mu}$, with 
the components
\be
\xi_t={\cal T}(t,r)Y_{l} (\theta)\,,\qquad 
\xi_r={\cal R}(t,r)Y_{l} (\theta)\,,\qquad 
\xi_{\theta}= \Theta (t,r) Y_{l,\theta} (\theta)\,,
\qquad 
\xi_{\varphi}=-\Lambda (t,r) (\sin \theta) 
Y_{l,\theta} (\theta)\,.
\ee
Note that $\Lambda$ is associated with the odd modes, 
while ${\cal T}$, ${\cal R}$, and $\Theta $ are 
related to the even modes. 
Metric perturbations transform as 
\be
Q \to Q+\dot{\Lambda}\,,\qquad 
W \to W+\Lambda'-\frac{2}{r} \Lambda\,,\qquad
U \to U+2 \Lambda\,,
\label{gaugeodd}
\ee
in the odd-parity sector, and 
\ba
& &
H_0 \to H_0+\frac{2}{f} \dot{{\cal T}}
-\frac{f' h}{f}{\cal R}\,,\qquad 
H_1 \to H_1+\dot{{\cal R}}+{\cal T}'
-\frac{f'}{f}{\cal T}\,,\qquad
H_2 \to H_2+2h{\cal R}'+h' {\cal R}\,,
\label{H1tra} 
\nonumber \\
& &
h_0 \to h_0+{\cal T}+\dot{\Theta}\,,\qquad 
h_1 \to h_1+{\cal R}+\Theta'
-\frac{2}{r}\Theta\,,\qquad
K \to K+\frac{2}{r}h{\cal R}\,,\qquad 
G \to G+\frac{2}{r^2}\Theta\,,\label{beal}
\ea
in the even-parity sector.

For $l \geq 2$, we can choose the gauge $U=0$ to fix $\Lambda$. In the same case, the possible gauge choice in the even-parity sector is $h_0=0$, $K=0$, and $G=0$, under which ${\cal T}$, ${\cal R}$, and $\Theta$ are fixed.

For $l=0$, all the odd-parity perturbations vanish identically. 
Moreover, the contributions to the action arising from even-parity perturbations $h_0$, $h_1$, $G$ also disappear, leaving $\Theta$ as a residual gauge degree of freedom (we can still set $\mathcal{R}$ to make $K$ vanish, and, by fixing appropriate boundary conditions at spatial infinity, we can choose $\mathcal{T}$ to set $H_1$ to zero, as in this case, the perturbation variables satisfy spherical symmetry and the metric can be brought to a diagonal form with the angular part to be $r^2{\rm d}\Omega^2$).

For $l=1$, the contributions to the action arising from the perturbation $U$ vanish, and hence we need to choose another gauge such as $W=0$ to fix $\Lambda$. 
Moreover, terms associated with $K$ and $G$ appear only as the combination $K-G$ and its derivatives in the action, so that there is also the residual gauge degree of freedom after imposing $K=G$. By fixing appropriate boundary conditions at spatial infinity, we can further choose $h_0=0$ and $h_1=0$ 
(or, $h_0=0$ and $H_2=0$) to set $\mathcal{R}$ and $\mathcal{T}$. 

In Secs.~\ref{eleBH} and \ref{magBH}, we will study the perturbations of purely electrically and magnetically charged BHs, respectively, by separating the analysis into three different cases: (1) $l \geq 2$, (2) $l=0$, and (3) $l=1$.

In the rest of this section, we expand the action (\ref{action}) up to quadratic order in perturbations by choosing the gauge
\be
U=0\,,\qquad h_0=0\,,\qquad K=0\,,
\qquad G=0\,.
\label{RWgauge}
\ee
The discussion of how to derive the reduced action of dynamical perturbations under the gauge choice (\ref{RWgauge}) is valid for $l \geq 2$.
Integrating the second-order action of perturbations with respect to $\theta$ and $\varphi$, performing the integration by parts, and dropping some boundary terms, the resulting action is expressed in the form ${\cal S}^{(2)}=\int \rd t \rd r\,{\cal L}$, where 
\be
{\cal L}={\cal L}_A+{\cal L}_B\,,
\label{Ltotal}
\ee
with
\ba
{\cal L}_A &=& 
L \biggl[ p_1 \left( \dot{W}-Q'+\frac{2Q}{r} \right)^2 
+ \left\{ p_2 \delta A + p_3 \left( \delta A_0 +\frac{q_M Q}{r^2} 
\right)+ p_4 \left( \delta A' + \frac{q_M}{r^2}h_1 \right) 
\right\} \left( \dot{W}-Q'+\frac{2Q}{r} \right) \nonumber \\
& &\quad+ p_5\dot{\delta A}^2 + p_6 \delta A'^2 + p_7 \delta A^2 
+ p_8 W^2 + p_9 Q^2 + p_{10}Q \delta A + p_{11} Q \delta A_0
+ p_{12} Q h_1 + p_{13} W \delta A_1 \biggr]\,,
\label{LA}
\ea
and 
\ba
{\cal L}_B &=& 
a_0 H_0^2 + H_0 \left[ a_1 H_2' + L a_2 h_1' + (a_3+L a_4) H_2 
+ L a_5 h_1+ L a_6 \delta A' + L a_7 \delta A \right] \nonumber \\
& &+ L b_0 H_1^2 + H_1 \left( b_1 \dot{H}_2 + L b_2 \dot{h}_1 
+ L b_3 \delta A_1 +L b_4 W+ L b_5 \dot{\delta A} \right) 
+ c_0 H_2^2 
+ H_2 (L c_1 h_1+ L c_2 \delta A_0  
+L c_3 Q+L c_4 \delta A) \nonumber \\
& &
+ L d_0 \dot{h}_1^2 + L d_1 h_1^2 
+ L \dot{h}_1 (d_2 \delta A_1+d_3 W + d_4 \dot{\delta A}) 
+ L h_1 ( d_5 \delta A_0  
+ d_6 \delta A') \nonumber \\
& &
+ s_1 (\delta A_0'-\dot{\delta A}_1)^2 
+\left( s_2 H_0+s_3 H_2+L s_4 h_1 \right)(\delta A_0'-\dot{\delta A}_1)
+ L s_5 \delta{A}_0^2 
+ L s_6 \delta A_1^2\,.
\label{LB}
\ea
Here, a dot represents a derivative with respect to $t$, and 
\be
L \equiv l(l+1)\,.
\ee
The $r$-dependent coefficients $p_1$, $a_0$, etc.\ are presented in Appendix A.
In the presence of a nonvanishing magnetic charge $q_M$, the perturbations in the odd- and even-parity sectors mix. 
In this case, we need to deal with them at once to obtain the second-order Lagrangian of dynamical perturbations. 
To identify the dynamical degrees of freedom, we introduce two Lagrange multipliers $\chi_1$ and $V$, as\footnote{The coefficients in the definition of $\chi_1$ have been chosen to set all the terms in $\dot{W}^2$ and $\dot{W}$ to vanish, except the one in $\dot{W}\chi_1$.}
\ba
{\cal L}_2 &=& 
{\cal L}_A+{\cal L}_B-L p_1 \left[ \dot{W}-Q'+\frac{2Q}{r} 
+\frac{1}{2p_1} \left\{ p_2 \delta A+p_3 
\left( \delta A_0+\frac{q_M}{r^2}Q \right)
+p_4 \left( \delta A'+\frac{q_M}{r^2}h_1 \right) \right\}
-\chi_1 \right]^2 \nonumber \\
& &\qquad\qquad\quad 
-s_1 \left[ \delta A_0'-\dot{\delta A}_1+\frac{1}{2s_1} 
\left( s_2 H_0+s_3 H_2+L s_4 h_1 \right)-V \right]^2\,.
\label{L2}
\ea
Varying Eq.~(\ref{L2}) with respect to $\chi_1$ and $V$, respectively, we obtain
\ba
\hspace{-0.5cm}
\chi_1 &=&
\dot{W}-Q'+\frac{2Q}{r} 
+\frac{1}{2p_1} \left[ p_2 \delta A+p_3 
\left( \delta A_0+\frac{q_M}{r^2}Q \right)
+p_4 \left( \delta A'+\frac{q_M}{r^2}h_1 \right) \right]
\nonumber \\
\hspace{-0.5cm}
&=& \dot{W}-Q'+\frac{2Q}{r}
-\frac{2r}{\Mpl^2 r^4+4\beta q_M^2} 
\left[ r(r^2-8 \beta h )A_0' \delta A 
-12 \beta q_M \left( \delta A_0 + \frac{q_M}{r^2}Q \right) 
+4 \beta h A_0' (r^2 \delta A' + q_M h_1) \right],
\label{chi1} \\
\hspace{-0.5cm}
V &=& \delta A_0'-\dot{\delta A}_1+\frac{1}{2s_1} 
\left( s_2 H_0+s_3 H_2+L s_4 h_1 \right)\nonumber \\
\hspace{-0.3cm}
&=& \delta A_0'-\dot{\delta A}_1+\frac{A_0'}{2} \left[ 
H_0-\frac{r^2-8 \beta (3h-1)}{r^2-8 \beta (h-1)}H_2
-\frac{16 L \beta h}{r\{ r^2-8 \beta (h-1)\}}h_1 \right]\,.
\label{V}
\ea
On using these relations, it follows that the Lagrangian ${\cal L}_2$ is equivalent to ${\cal L}$. The field $V$ corresponds to the vector-field perturbation in the even-parity sector. 
For $q_M=0$ or $\beta=0$, the field $\chi_1$ is composed of only the odd-parity perturbations $W$, $Q$, $\delta A$ and their derivatives. In these limits, $\chi_1$ corresponds to the gravitational perturbation in the odd-parity sector. 
For $q_M \neq 0$ and $\beta \neq 0$, $\chi_1$ acquires the contributions of even-parity perturbations $\delta A_0$ and $h_1$. 

Varying Eq.~(\ref{L2}) with respect to $Q$ and $W$, we obtain their field equations of motion, respectively. 
We use them to remove $Q$, $W$, and their derivatives from the Lagrangian ${\cal L}_2$.
After this procedure, we also eliminate the fields $\delta A_0$ and $\delta A_1$ from ${\cal L}_2$ by using their equations of motion.
Since there is the relation $a_2=-a_1/r$, the combination $H_0(a_1 H_2' + L a_2 h_1')$ appearing in Eq.~(\ref{LB}) is equivalent to $a_1 H_0 (H_2'-L h_1'/r)$. 
To express the radial derivatives $H_2'$ and $h_1'$ in terms of a single perturbed quantity and its radial derivative, we introduce the following field
\be
\chi_2=H_2-\frac{L}{r}h_1\,.
\label{chi2}
\ee
Then, we substitute $H_2=\chi_2+L h_1/r$ and their derivatives 
into the Lagrangian to eliminate the $H_2$-dependent terms. 

To integrate out the field $H_0$ from ${\cal L}_2$, we first exploit the following relation 
\be
a_0=\frac{s_2^2}{4s_1}\,,
\label{a0re}
\ee
with which the coefficient in front of $H_0^2$ vanishes. 
Then, the Lagrangian contains only the terms linear in $H_0$, 
whose equation of motion gives a constraint on $h_1$.
This latter equation is used to remove the field $h_1$ and 
its derivatives from the Lagrangian.
Finally, we vary ${\cal L}_2$ with respect to $H_1$ and solve 
it for $H_1$. On using this latter relation, we end up with the Lagrangian ${\cal L}_2$ containing four dynamical fields given by
\be
\vec{\Psi}^t= \left( \chi_1, \delta A, \chi_2, V \right)\,,
\ee
as well as their $t$ and $r$ derivatives. 
Note that $\delta A$ corresponds to the vector-field perturbation in the odd-parity sector, whereas $\chi_2$ to the gravitational perturbation in the even-parity sector. 
Since the Lagrangian of four dynamical perturbations is cumbersome for the dyon BH with mixed electric and magnetic charges, we will not show its explicit form. 
Instead, in subsequent sections, we will consider the two cases: (i) $q_M=0$ and (ii) $q_E=0$, in turn.

\section{Linear stability of electrically charged BHs}
\label{eleBH}

In this section, we derive the linear stability conditions of purely electrically charged BHs characterized by 
\be
q_E \neq 0\,,\qquad q_M=0\,.
\ee
We will study the three different cases: 
(1) $l \geq 2$, (2) $l=0$, and (3) $l=1$, in turn.

\subsection{$l \geq 2$}

Choosing the gauge (\ref{RWgauge}) for the multiples $l \geq 2$, the second-order perturbed action ${\cal L}$ is composed of the two contributions ${\cal L}_A$ and ${\cal L}_B$ given by Eqs.~(\ref{LA}) and (\ref{LB}), respectively.
For the electric BH, the coefficients in ${\cal L}$ have the following properties
\be
p_3=p_{11}=p_{12}=p_{13}=0\,,\qquad
a_6=a_7=0\,,\qquad
b_4=b_5=0\,,\qquad
c_3=c_4=0\,,\qquad
d_3=d_4=d_6=0\,.
\ee
Then, we find that ${\cal L}_A$ consists of only the odd-parity perturbations $Q$, $W$, $\delta A$ and their derivatives, while ${\cal L}_B$ contains the even-parity perturbations $H_0$, $H_1$, $H_2$, $h_1$, $\delta A_0$, $\delta A_1$ and their derivatives alone. 
In the following, we will discuss the linear stability of BHs in the odd- and even-parity sectors separately.

\subsubsection{Odd-parity perturbations}

In the odd-parity sector, we consider the following second-order Lagrangian  
\ba
{\cal L}_{\rm odd}
&=&
L \biggl[ p_1 \left( \dot{W}-Q'+\frac{2Q}{r} \right)^2 
+ \left( p_2 \delta A + p_4 \delta A'
\right) \left( \dot{W}-Q'+\frac{2Q}{r} \right)\nonumber \\
& &\quad
+ p_5\dot{\delta A}^2 + p_6 \delta A'^2 + p_7 \delta A^2 
+ p_8 W^2 + p_9 Q^2 + p_{10}Q \delta A  \biggr]\nonumber \\
& &
-L p_1 \left[ \dot{W}-Q'+\frac{2Q}{r} 
+\frac{1}{2p_1} \left( p_2 \delta A
+p_4 \delta A' \right)
-\chi_1 \right]^2\,.
\label{Lodd}
\ea
Varying Eq.~(\ref{Lodd}) with respect to $\chi_1$, we have
\be
\chi_1=\dot{W}-Q'+\frac{2Q}{r} 
+\frac{1}{2p_1} \left( p_2 \delta A
+p_4 \delta A' \right)\,,
\ee
and hence ${\cal L}_{\rm odd}$ is equivalent to 
${\cal L}_A$. 
Varying ${\cal L}_{\rm odd}$ with respect to $W$ and $Q$ and solving the equations of motion 
for these fields, we obtain 
\ba
& &
W=\frac{p_1}{p_8} \dot{\chi}_1\,,\\
& &
Q=-\frac{2r p_1 \chi_1'+2(rp_1'+2p_1)\chi_1
+p_{10}r \delta A}{2rp_9}\,,
\ea
which are valid for $p_8 \neq 0$ and $p_9 \neq 0$.
We use these relations to eliminate the terms $\dot{W}$, $Q'$, and $Q$ in Eq.~(\ref{Lodd}). 
After several integrations by parts, the second-order Lagrangian reduces to the form 
\be
{\cal L}_{\rm odd}= 
L \left( \dot{\vec{\mathcal{X}}}^{t}{\bm K}\dot{\vec{\mathcal{X}}}
+\vec{\mathcal{X}}'^{t}{\bm G}\vec{\mathcal{X}}'
+\vec{\mathcal{X}}^{t}{\bm M}\vec{\mathcal{X}}
+\vec{\mathcal{X}}'^{t}{\bm S}\vec{\mathcal{X}}\right)
\,,\label{Lodd2}
\ee
where ${\bm K}, {\bm G}, {\bm M}$ are $2 \times 2$ symmetric matrices, ${\bm S}$ is a $2 \times 2$ antisymmetric matrix, and 
\be
\vec{\mathcal{X}}^{t}=\left(\chi_1,\delta A \right)\,.
\label{mathcalX}
\ee
The nonvanishing components of ${\bm K}, {\bm G}, {\bm M}, {\bm S}$ are 
\ba
& &
K_{11}=-\frac{p_1^2}{p_8}\,,\qquad K_{22}=p_5\,,
\qquad G_{11}=-\frac{p_1^2}{p_9}\,,\qquad
G_{22}=p_6-\frac{p_4^2}{4p_1}\,,\nonumber \\
& &
M_{11}=-p_1-\frac{p_1 [r p_9'(r p_1'+2p_1)
+p_9(6p_1-r^2p_1'')]}{r^2 p_9^2}\,,\nonumber \\
& &
M_{22}=p_7-\frac{p_{10}^2}{4p_9}
-\frac{p_1' p_2 p_4+p_1(p_2^2-p_2' p_4-p_2 p_4')}{4p_1^2}\,,
\nonumber \\
& &
M_{12}=\frac{1}{4} (2p_2-p_4')
+\frac{p_9[rp_1 p_{10}'-p_{10}(rp_1'+4p_1)]
-rp_1 p_9' p_{10}}{4rp_9^2}\,,\nonumber \\
& &
S_{12}=-S_{21}=-\frac{1}{2} \left( p_4
+\frac{p_1 p_{10}}{p_9} \right)\,.
\ea
Thus, there are two dynamical perturbations $\chi_1$ and $\delta A$ in the odd-parity sector. 
The ghosts are absent under the two conditions $K_{11}=-p_1^2/p_8>0$ and $K_{22}=p_5>0$, which translate to
\ba
-p_8 &=& 
\frac{\sqrt{h} (\Mpl^2 f-4\beta h A_0'^2)}
{4r^2 \sqrt{f}}(L-2)>0\,,\\
p_5 &=& \frac{r-4\beta h'}{2r\sqrt{fh}}>0\,.
\label{p5con}
\ea
For the multipoles $l \geq 2$, these inequalities hold if
\ba
\frac{r+8 \beta h'}{r^2-8\beta (h-1)}
&>&0\,,
\label{nogo1}\\
r-4\beta h' &>& 0\,,
\label{nogo2}
\ea
where we used Eq.~(\ref{back1}) with Eq.~(\ref{back4}) to eliminate $q_E$. These two conditions are satisfied for $\beta$ close to 0. 

Varying the Lagrangian (\ref{Lodd2}) with respect to $\chi_1$ and $\delta A$, the resulting perturbation equations of motion are given, respectively, by 
\ba
& &
K_{11} \ddot{\chi}_1+G_{11} \chi_1''+G_{11}' \chi_1'
-M_{11}\chi_1+S_{12} \delta A' 
-\left( M_{12}-\frac{1}{2}S_{12}' \right)\delta A=0\,,
\label{perodd1}\\
& &
K_{22} \ddot{\delta A}+G_{22} \delta A''+G_{22}'\delta A'
-M_{22} \delta A-S_{12} \chi_1'
-\left( M_{12}+\frac{1}{2}S_{12}' \right)\chi_1=0\,.
\label{perodd2}
\ea
We derive the propagation speeds of $\chi_1$ and $\delta A$ by assuming the solutions to Eqs.~(\ref{perodd1}) and (\ref{perodd2}) in the form $\vec{\mathcal{X}}^t=\vec{\mathcal{X}}_0^{t} e^{i (\omega t-kr)}$, 
where $\vec{\mathcal{X}}_0^{t}$ is a constant vector. 
For the radial propagation, we take the limits of large frequencies $\omega$ and momenta $k$ in Eqs.~(\ref{perodd1}) and (\ref{perodd2}).
Then, we obtain the following two dispersion relations 
\be
\omega^2=-\frac{G_{11}}{K_{11}}k^2\,,\qquad 
\omega^2=-\frac{G_{22}}{K_{22}}k^2\,.
\label{disodd}
\ee
We define the radial propagation speed $c_r$ in terms of the proper time $\tau=\int \sqrt{f}\,\rd t$ and the rescaled radial coordinate $\tilde{r}=\int \rd r/\sqrt{h}$, as $c_r=\rd \tilde{r}/\rd \tau=\hat{c}_r/\sqrt{fh}$, where $\hat{c}_r=\rd r/\rd t=\omega/k$.
On using Eq.~(\ref{disodd}), the two squared propagation speeds are given by 
\ba
c_{r1}^2 &=& 
-\frac{1}{fh}\frac{G_{11}}{K_{11}}
=-\frac{1}{fh}\frac{p_8}{p_9}=1\,,\\
c_{r2}^2 &=& 
-\frac{1}{fh}\frac{G_{22}}{K_{22}}
=\frac{1}{fh}\frac{p_4^2-4p_1 p_6}{4p_1 p_5}=1\,.
\label{crodd}
\ea
Hence the two fields $\chi_1$ and $\delta A$ propagate with the speed of light along the radial direction. 

For the angular propagation, we take the limits of large $\omega^2$ and $L \gg 1$ in Eqs.~(\ref{perodd1}) and (\ref{perodd2}) and ignore the radial derivative terms for perturbations.
To allow for the existence of nonvanishing solutions to $\vec{\mathcal{X}}_0^t$, we require that 
\be
\left( \omega^2 K_{11}+M_{11} \right)
\left( \omega^2 K_{22}+M_{22} \right)
-M_{12}^2+\frac{S_{12}'^2}{4}=0\,.
\label{disan}
\ee
The matrix components have the multipole dependences $K_{11} \propto L^{-1}$, $M_{11} \propto L^{0}$, $K_{22} \propto L^{0}$, $M_{22} \propto L$, $M_{12} \propto L^{0}$, $S_{12}' \propto L^{0}$, respectively. 
The angular propagation speed in proper time is defined by $c_{\Omega}=r \rd \theta/\rd \tau=\hat{c}_{\Omega}/\sqrt{f}$, where $\hat{c}_{\Omega}=r\rd \theta/\rd t$ satisfies $\omega^2=\hat{c}_{\Omega}^2 l^2/r^2$. 
Then, we look for solutions with the dispersion relation $\omega^2={\cal O}(1) L \gg 1$ in Eq.~(\ref{disan}). 
In this case, the leading-order solutions to Eq.~(\ref{disan}) 
are given by $\omega^2=-M_{11}/K_{11}$ and $\omega^2=-M_{22}/K_{22}$.
As a result, we obtain the following squared angular propagation speeds
\ba
c_{\Omega 1}^2 &=& 
-\frac{r^2}{fL} \frac{M_{11}}{K_{11}} \biggr|_{L \to \infty}
=1-\frac{4\beta h A_0'^2}{\Mpl^2 f}
=\frac{r(r+8 \beta h')}{r^2 - 8\beta(h - 1)}
\,,\label{cO1}\\
c_{\Omega 2}^2 &=& 
-\frac{r^2}{fL} \frac{M_{22}}{K_{22}} \biggr|_{L \to \infty} 
\nonumber \\
&=& [r^6 + 8 \beta (3 rh' - h + 1)r^4 + 512 \beta^3 h'^2 (rh' - 3h + 3)r^2  
+ 192 \beta^2 \{ h(3-4r h' )-3h^2  
+ r h' (rh' + 1) \}r^2 \nonumber \\
& &- 4096 \beta^4 h'^3 (h - 1)r]
/[(r-4\beta h')(r+8\beta h' ) 
\{ r^2 - 8 \beta  (h - 1)\}^2]\,.
\label{cO2}
\ea
So long as the no-ghost condition (\ref{nogo1}) is satisfied, we have that $c_{\Omega 1}^2>0$. 
The angular Laplacian stability for the field $\delta A$ requires that 
\be
c_{\Omega 2}^2>0\,.
\label{cO2con}
\ee
The expansion of $c_{\Omega 2}^2$ around 
$\beta=0$ gives 
\be
c_{\Omega 2}^2=1-\frac{12 \Mpl^2 r^2 (h - 1)
+10q_E^2}{\Mpl^2 r^4}\beta+{\cal O}(\beta^2)\,,
\ee
and hence $c_{\Omega 2}^2 \to 1$ as $\beta \to 0$. 
In the regime of asymptotic flatness ($h \to 1$, $h' \to 0$ as $r \to\infty$), both $c_{\Omega 1}^2$ and $c_{\Omega 2}^2$
approach 1.

To obtain the effective potentials of the dynamical fields $\chi_1$ and $\delta A$, we consider the solutions to Eqs.~(\ref{perodd1}) and (\ref{perodd2}) in the forms
\be
\chi_1={\cal C}_0(r) \tilde{\chi}_1 (r) e^{i \omega t}\,,\qquad 
\delta A={\cal D}_0(r) \tilde{\delta A} (r) e^{i \omega t}\,,
\ee
where ${\cal C}_0(r)$, $\tilde{\chi}_1 (r)$, ${\cal D}_0(r)$, $\tilde{\delta A} (r)$ are $r$-dependent functions. 
We choose ${\cal C}_0(r)$ and ${\cal D}_0(r)$ to eliminate the first derivatives $\rd \tilde{\chi}_1/\rd r_*$ and $\rd \tilde{\delta A}/\rd r_*$ in the differential equations for $\tilde{\chi}_1$ and $\tilde{\delta A}$ \cite{Blazquez-Salcedo:2018jnn,Antoniou:2022agj,Minamitsuji:2024twp}, 
where 
\be
r_*=\int g_s\,\rd r\,,\qquad {\rm with} 
\qquad g_s=\frac{1}{\sqrt{fh}}\,,
\ee
is the tortoise coordinate.
These choices amount to
\be
\frac{{\cal C}'_0}{{\cal C}_0}=\frac{1}{2}
\left( {\cal C}_1-\frac{g_s'}{g_s} \right)\,,\qquad
\frac{{\cal D}'_0}{{\cal D}_0}=\frac{1}{2}
\left( {\cal D}_1-\frac{g_s'}{g_s} \right)\,,
\ee
where 
\be
{\cal C}_1=-\frac{G_{11}'}{G_{11}}\,,\qquad 
{\cal D}_1=-\frac{G_{22}'}{G_{22}}\,.
\ee
Then, we obtain
\ba
& &
\left[ \frac{\rd^2}{\rd r_*^2}-V_1(r)+\omega^2 \right]
\tilde{\chi}_1=-\frac{fh}{{\cal C}_0 G_{11}} \left[ S_{12} 
\left( {\cal D}_0 \tilde{\delta A}'+ 
{\cal D}_0' \tilde{\delta A} \right)
-\left( M_{12}-\frac{1}{2} S_{12}' \right)
{\cal D}_0 \tilde{\delta A} \right]\,,\label{pereq1}\\
& &
\left[ \frac{\rd^2}{\rd r_*^2}-V_2(r)+\omega^2 \right]
\tilde{\delta A}=\frac{fh}{{\cal D}_0 G_{22}} 
\left[ S_{12} 
\left( {\cal C}_0 \tilde{\chi}'_1+ 
{\cal C}_0'\tilde{\chi}_1 \right)
+\left( M_{12}+\frac{1}{2} S_{12}' \right)
{\cal C}_0\tilde{\chi}_1 \right]\,,\label{pereq2}
\ea
where 
\ba
V_1(r) &=& \frac{1}{g_s^2} \left( \frac{M_{11}}{G_{11}}
+\frac{{\cal C}_1^2}{4}-\frac{{\cal C}_1'}{2}
-\frac{3g_s'^2}{4g_s^2}+\frac{g_s''}{2g_s} \right)\,,
\label{V1}\\
V_2(r) &=& \frac{1}{g_s^2} \left( \frac{M_{22}}{G_{22}}
+\frac{{\cal D}_1^2}{4}-\frac{{\cal D}_1'}{2}
-\frac{3g_s'^2}{4g_s^2}+\frac{g_s''}{2g_s} \right)\,.
\label{V2}
\ea
The two effective potentials $V_1(r)$ and $V_2(r)$ determine the radial tachyonic stability of BHs. 
For small $\beta$ close to 0, the expansions of $V_1(r)$ and $V_2(r)$ around $\beta=0$ lead to 
\ba
\hspace{-0.7cm}
V_1(r) &=& \frac{f [2 \Mpl^2 r^2 (L + 3h - 3) 
+ q_E^2]}{2\Mpl^2 r^4}-\frac{4q_E^2 f
[\Mpl^2 r^2 (L - 16h + 1) - q_E^2]}{\Mpl^4 r^8}\beta
+{\cal O}(\beta^2)\,,\label{V1ap}\\
\hspace{-0.7cm}
V_2(r) &=& \frac{f (L \Mpl^2 r^2 + 2 q_E^2)}{\Mpl^2 r^4}
+\frac{f [5 q_E^4 -4 (5L-38 h + 20 ) \Mpl^2 q_E^2 r^2 
-12 (h - 1) (2L-5h+1 ) \Mpl^4 r^4]}
{2 \Mpl^4 r^8}\beta+{\cal O}(\beta^2)\,.
\label{V2ap}
\ea
The first terms in Eqs.~(\ref{V1ap}) and (\ref{V2ap}) correspond to the effective potentials of the RN BH in GR \cite{Moncrief:1974gw, Moncrief:1974ng, Moncrief:1975sb, Zerilli:1974ai} for the gravitational and electromagnetic perturbations, 
respectively. The coupling $\beta$ gives rise to corrections to 
both $V_1(r)$ and $V_2(r)$, which should also affect the quasinormal modes of BHs.

If we take the limit $L \gg 1$ in Eqs.~(\ref{V1}) 
and (\ref{V2}) without using the approximation of the small coupling constant $\beta$, it follows that 
\be
V_1(r)=\frac{fL}{r^2}c_{\Omega 1}^2\,,\qquad 
V_2(r)=\frac{fL}{r^2}c_{\Omega 2}^2\,,\qquad 
\label{V12}
\ee
where $c_{\Omega 1}^2$ and $c_{\Omega 2}^2$ are given by Eqs.~(\ref{cO1}) and (\ref{cO2}), respectively. 
In this limit, the right-hand sides of Eqs.~(\ref{pereq1}) and (\ref{pereq2}) can be neglected because the mixing coefficients $M_{12}$ and $S_{12}$ are suppressed.
Hence, for $L \gg 1$, the odd-parity dynamics is described by a decoupled system of the two dynamical perturbations $\tilde{\chi}_1$ and $\tilde{\delta A}$.
In the same eikonal limit, the deviations of $c_{\Omega 1}^2$ and $c_{\Omega 2}^2$ from 1 induced by the coupling $\beta$ modify the shapes of the effective potentials in GR. Furthermore, the fact that $c_{\Omega 1}^2\ne c_{\Omega 2}^2$ implies that the peak of at least one of the potentials $V_1(r)$ and $V_2(r)$ in the eikonal limit is not located at the photon sphere \cite{Cardoso:2008bp}. The eikonal correspondence is thus broken at least for one of the modes. 
The breaking of the eikonal correspondence has been found in some modified theories of gravity, such as those with nonminimal matter-gravity couplings \cite{Chen:2018vuw, Glampedakis:2019dqh, Chen:2019dip, Chen:2021cts} or those with higher spacetime dimensions \cite{Konoplya:2017wot, Moura:2021eln}. 
Therefore, observationally testing the correspondence could be 
a potential way of searching physics beyond GR \cite{Chen:2022nlw}.

\subsubsection{Even-parity perturbations}

In the even-parity sector, we consider the following second-order Lagrangian 
\ba
{\cal L}_{\rm even} &=& 
a_0 H_0^2 + H_0 \left[ a_1 H_2' + L a_2 h_1' + (a_3+L a_4) H_2 
+ L a_5 h_1  \right] 
+ L b_0 H_1^2 + H_1 \left( b_1 \dot{H}_2 + L b_2 \dot{h}_1 
+ L b_3 \delta A_1 \right)  \nonumber \\
& &
+ c_0 H_2^2 
+ L H_2 \left( c_1 h_1+ c_2 \delta A_0 \right) 
+ L \left( d_0 \dot{h}_1^2 + d_1 h_1^2 
+ d_2 \dot{h}_1 \delta A_1 
+ d_5 h_1 \delta A_0 \right)  \nonumber \\
& &
+ s_1 \left( \delta A_0'-\dot{\delta A}_1 \right)^2 
+\left( s_2 H_0+s_3 H_2+L s_4 h_1 \right)
\left( \delta A_0'-\dot{\delta A}_1 \right)
+ L s_5 \delta{A}_0^2 
+ L s_6 \delta A_1^2 \nonumber \\
& &
-s_1 \left[ \delta A_0'-\dot{\delta A}_1+\frac{1}{2s_1} 
\left( s_2 H_0+s_3 H_2+L s_4 h_1 \right)-V \right]^2\,,
\label{Leven}
\ea
whose variation with respect to $V$ leads to Eq.~(\ref{V}). 
Varying ${\cal L}_{\rm even}$ with respect to $\delta A_0$ and $\delta A_1$ and solving the perturbation equations for $\delta A_0$ and $\delta A_1$, we obtain 
\ba
\delta A_0 &=& \frac{2s_1' V+2s_1 V'-Lc_2 H_2-L d_5 h_1}{2L s_5}\,,\\
\delta A_1 &=&-\frac{2s_1 \dot{V}+L b_3 H_1+L d_2 
\dot{h}_1}{2L s_6}\,,
\ea
which are valid for $s_5 \neq 0$ and $s_6 \neq 0$. 
We use these relations to eliminate the terms $\delta A_0$, $\delta A_1$, and their derivatives from Eq.~(\ref{Leven}).
We introduce the dynamical field $\chi_2$ defined by Eq.~(\ref{chi2}) and remove $H_2$ and its derivatives from the Lagrangian.
On using the relation (\ref{a0re}) and varying ${\cal L}_{\rm even}$ with respect to $H_0$, we can express $h_1$ in terms of $\chi_2'$, $\chi_2$, and $V$. Then, we can eliminate the $h_1$-dependent terms in ${\cal L}_{\rm even}$. 
Finally, we remove the $H_1$-dependent terms by using its equation of motion. 
After the integration by parts, the resulting second-order action is expressed in the form 
\be
{\cal L}_{\rm even}= 
\dot{\vec{\mathcal{Y}}}^{t} \tilde{{\bm K}} 
\dot{\vec{\mathcal{Y}}}
+\vec{\mathcal{Y}}'^{t} \tilde{{\bm G}}\vec{\mathcal{Y}}'
+\vec{\mathcal{Y}}^{t} \tilde{{\bm M}}\vec{\mathcal{Y}}
+\vec{\mathcal{Y}}'^{t} \tilde{{\bm S}}\vec{\mathcal{Y}}\,,
\label{Leven2}
\ee
where $\tilde{{\bm K}}, \tilde{{\bm G}}, \tilde{{\bm M}}$ are $2 \times 2$ symmetric matrices, $\tilde{{\bm S}}$ is a $2 \times 2$ antisymmetric matrix with the components $\tilde{S}_{11}=\tilde{S}_{22}=0$ and $\tilde{S}_{12}=-\tilde{S}_{21} \neq 0$, and 
\be
\vec{\mathcal{Y}}^{t}=\left(\chi_2, V \right)\,.
\label{mathcalX}
\ee
Unlike odd-parity perturbations, there are nonvanishing off-diagonal components $\tilde{K}_{12}=\tilde{K}_{21}$ and $\tilde{G}_{12}=\tilde{G}_{21}$ in $\tilde{\bm K}$ and $\tilde{\bm G}$, respectively.

Varying the Lagrangian (\ref{Leven2}) with respect to $\chi_2$ and $V$, the resulting perturbation equations of motion are 
\ba
& &
\tilde{K}_{11} \ddot{\chi}_2 + \tilde{K}_{12} 
\ddot{V}+\tilde{G}_{11} \chi_2'' 
+ \tilde{G}_{12}V'' + \tilde{G}_{11}'\chi_2' 
-\tilde{M}_{11} \chi_2
+ \left( \tilde{G}_{12}'+\tilde{S}_{12} \right)V' 
+ \left( \tilde{S}_{12}/2 - \tilde{M}_{12} 
\right) V=0\,,\label{pereven1}\\
& &
\tilde{K}_{22} \ddot{V} + \tilde{K}_{12} \ddot{\chi}_2 
+ \tilde{G}_{22} V'' + \tilde{G}_{12} \chi_2''
+\tilde{G}_{22}' V'- \tilde{M}_{22} V 
+ \left( \tilde{G}_{12}' - \tilde{S}_{12} \right)\chi_2' 
-\left(\tilde{S}_{12}'/2+\tilde{M}_{12} \right) \chi_2=0\,.
\label{pereven2}
\ea
The absence of ghosts requires the following two conditions 
\ba
\tilde{K}_{11}\tilde{K}_{22}-\tilde{K}_{12}^2
&=&\frac{(L-2) \Mpl^2 r^4 h^2 [r^2-8 \beta (h-1)](r+8 \beta h')}
{2 L^2 f^2 (r-4\beta h')(L+rh'-2h)^2}>0\,,\label{nogo3}\\
\tilde{K}_{22} &=& 
\frac{r \sqrt{h} [r^2-8 \beta (h-1)]^2}
{2L f^{3/2} (r-4\beta h')}>0\,,
\ea
where we used Eq.~(\ref{back1}) with Eq.~(\ref{back4}). 
For the multipoles $l \ge 2$, these conditions are satisfied if
\ba
[r^2-8 \beta (h-1)](r+8 \beta h') &>&0\,,\label{nogo2c}\\
r-4\beta h' &>& 0\,,\label{nogo2d}
\ea
which are equivalent to the no-ghost conditions (\ref{nogo1}) and (\ref{nogo2}) derived for odd-parity perturbations. 
We also note that, under the condition (\ref{nogo3}), Eqs.~(\ref{pereven1}) and (\ref{pereven2}) can be solved for $\ddot{\chi}_2$ and $\ddot{V}$.

The radial propagation speeds can be found by assuming the solutions in the form $\vec{\mathcal{Y}}^t=\vec{\mathcal{Y}}_0^{t} e^{i (\omega t-kr)}$ and taking the large $\omega, k$ limits. 
In this regime, the first four terms in Eqs.~(\ref{pereven1}) and (\ref{pereven2}) are the dominant contributions to the perturbation equations of motion. 
The propagation speeds $c_r$ along the radial 
direction are known by solving the following equation 
\be
{\rm det}\,\left( fh c_r^2 \tilde{{\bm K}}+\tilde{{\bm G}} 
\right)=0\,,
\ee
which reduces to 
\be
\frac{ 2 \Mpl^6 (L - 2) r^4 h^4 [r^2 - 8\beta (h - 1)]^3 
[\Mpl^2 \{ r^2 - 8 \beta (h - 1) \}^2-4 \beta q_E^2 ] (c_r^2 - 1)^2}
{L^2[ 2(L - 3h + 1) \Mpl^2 \{ r^2 - 8 \beta (h - 1)\} -qE^2]^2
[2 \beta q_E^2 + \Mpl^2 \{ r^2 - 8(h - 1) \beta\} 
\{r^2+4 \beta (h - 1) \}]}=0\,.
\ee
Then, we obtain the two solutions 
\be
c_{r3}^2=1\,,\qquad 
c_{r4}^2=1\,,
\ee
and hence the two dynamical degrees of freedom $\chi_2$ and $V$ propagate with the speed of light 
in the radial direction. 

In the limit $L \gg 1$, the components of $\tilde{{\bm K}}$ and $\tilde{{\bm M}}$ contribute to the dispersion relation, while the components of $\tilde{{\bm S}}$ can be neglected. 
The angular propagation speeds $c_{\Omega}$ can be obtained by solving 
\be
{\rm det}\,\left( fL c_{\Omega}^2 \tilde{{\bm K}}
+r^2 \tilde{{\bm M}} 
\right)=0\,.
\label{detan}
\ee
Taking the limit $L \to \infty$, the solutions to Eq.~(\ref{detan}) are given by 
\ba
c_{\Omega 3}^2 &=& \frac{r (r-4 \beta h')}{r^2 - 8\beta(h - 1)}\,,
\label{cO3}\\
c_{\Omega 4}^2 &=& 
[r^6 + 4\beta (3r h' - 2h + 2)r^4 -96 \beta^2 
\{  h (rh' + 2) + rh' (2rh' - 3)-2h^2 \} r^2
+2048 \beta^4 h'^3 (h-1) r  \nonumber \\
& &
+256 \beta^3 \{ 6h(h - 1)^2 + 6(h - 1)^2 h'r - 6h'^2r^2 - h'^3 r^3 \}]
/[(r-4\beta h')(r+8\beta h' ) 
\{ r^2 - 8 \beta  (h - 1)\}^2]\,, 
\label{cO4}
\ea
so that $c_{\Omega 3}^2 \neq c_{\Omega 4}^2$.
This is again a manifestation that the eikonal correspondence is broken in this theory. 
We also note that all of the four angular propagation speeds in the odd- and even-parity sectors are different from each other.
In the limit $\beta \to 0$, both $c_{\Omega 3}^2$ and $c_{\Omega 4}^2$ reduce to 1. 
At spatial infinity, they also approach 1.

{}From the perturbation equations of motion, we can also derive the angular propagation speeds in the following way. 
Substituting $\chi_2=\tilde{\chi}_2(r) e^{i \omega t}$ and $V=\tilde{V}(r) e^{i \omega t}$ into Eqs.~(\ref{pereven1}) and (\ref{pereven2}) and solving them for $\tilde{\chi}_2(r)$ and $\tilde{V}(r)$, we find
\ba
& &
\tilde{\chi}_2''+g_s^2\mu_1 \tilde{\chi}_2'
+g_s^2(\omega^2+\mu_2) \tilde{\chi}_2+g_s^2 \left( 
\mu_3 \tilde{V}'+\mu_4 \tilde{V} \right)=0\,,
\label{perre1}\\
& &
\tilde{V}''+g_s^2\nu_1 \tilde{V}'
+g_s^2(\omega^2+\nu_2) \tilde{V}
+g_s^2 \left( \nu_3 \tilde{\chi}_2'
+\nu_4 \tilde{\chi}_2 \right)=0\,,
\label{perre2}
\ea
where $g_s=1/\sqrt{fh}$ and the $r$-dependent functions $\mu_{1,2,3,4}$ and $\nu_{1,2,3,4}$ are given in Appendix B. 
In the regime characterized by $\omega^2 \approx {\cal O}(1) L 
\gg (k r_h)^2 \gg 1$, we can ignore the radial derivatives of $\tilde{\chi}_2$ and $\tilde{V}$ in Eqs.~(\ref{perre1}) and (\ref{perre2}), so that 
\ba
& &
(\omega^2+\mu_2) \tilde{\chi}_2+\mu_4 \tilde{V} \simeq 0\,,\\
& &
\nu_4 \tilde{\chi}_2+(\omega^2+\nu_2) \tilde{V} \simeq 0\,.
\ea
To allow for the existence of nonvanishing solutions to $\tilde{\chi}_2$ and $\tilde{V}$, we require that 
\be
(\omega^2+\mu_2)(\omega^2+\nu_2)-\mu_4 \nu_4=0\,.
\label{diseven}
\ee
The angular propagation speeds $c_{\Omega}$ can be obtained by substituting $\omega^2=c_{\Omega}^2fL/r^2$ into Eq.~(\ref{diseven}). This procedure leads to the same values of $c_{\Omega 3}^2$ and $c_{\Omega 4}^2$ as those given in Eqs.~(\ref{cO3}) and (\ref{cO4}). 
Analogous to Eq.~(\ref{V12}), the quantities defined by
\be
V_3(r)=\frac{fL}{r^2}c_{\Omega 3}^2\,,\qquad 
V_4(r)=\frac{fL}{r^2}c_{\Omega 4}^2
\ee
can be interpreted as the effective potentials of even-parity perturbations in the eikonal limit.

\subsubsection{Summary of linear stability conditions}

Let us summarize the linear stability conditions derived for odd- and even-parity perturbations.
Under one of the no-ghost conditions given by Eq.~(\ref{nogo2d}), the Laplacian stability condition $c_{\Omega 3}^2>0$ is satisfied for $r^2-8 \beta (h-1)>0$.
To fulfill the other no-ghost condition (\ref{nogo2c}), we require that $r+8 \beta h'>0$. 
Under these inequalities, the Laplacian stability condition (\ref{cO2con}) associated with the odd-parity perturbation $\delta A$ translates to 
\ba
& &
r^6 + 8 \beta (3 rh' - h + 1)r^4 + 512 \beta^3 h'^2 
(rh' - 3h + 3)r^2  
+ 192 \beta^2 \{ h(3-4r h' )-3h^2  
+ r h' (rh' + 1) \}r^2 \nonumber \\
& &- 4096 \beta^4 h'^3 (h - 1)r>0\,.
\label{staodd}
\ea
Similarly, the Laplacian stability condition 
$c_{\Omega 4}^2>0$ holds if
\ba
& &
r^6 + 4\beta (3r h' - 2h + 2)r^4 -96 \beta^2 
\{  h (rh' + 2) + rh' (2rh' - 3)-2h^2 \} r^2
+2048 \beta^4 h'^3 (h-1) r  \nonumber \\
& &
+256 \beta^3 \{ 6h(h - 1)^2 + 6(h - 1)^2 h'r 
- 6h'^2r^2 - h'^3 r^3 \}>0\,.
\label{staeven}
\ea
In summary, the linear stability of electric 
BHs against odd- and even-parity perturbations 
is ensured for
\be
r-4 \beta h'>0\,,\qquad r+8 \beta h'>0\,,\qquad 
r^2-8 \beta (h-1)>0\,,\label{sta3}
\ee
and the inequalities (\ref{staodd}) and (\ref{staeven}). 
In the limit $\beta \to 0$, all of these conditions are trivially satisfied, with the luminal propagation of four dynamical perturbations along the radial and angular directions. For $\beta \neq 0$, the four propagation speeds in the angular direction are different from each other, while all of the radial propagation speeds are equivalent to 1. 
Since the deviation of linear stability conditions from GR is most significant at the outer horizon $r=r_h$, 
we only need to estimate those conditions at $r=r_h$.

\begin{figure}[ht]
\begin{center}
\includegraphics[height=2.8in,width=4.5in]{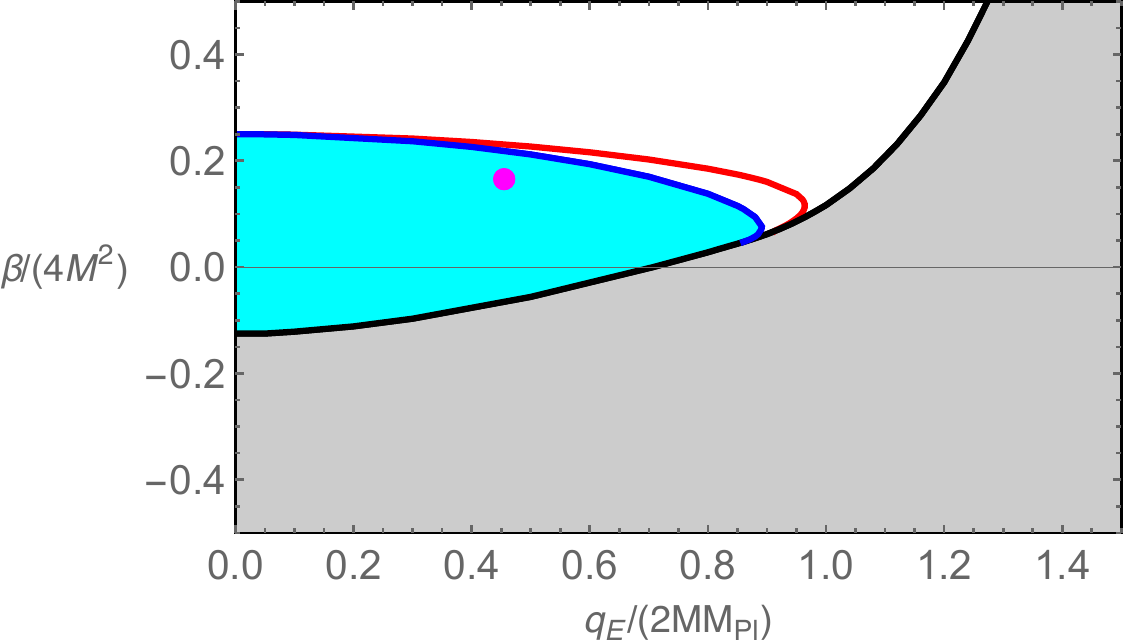}
\end{center}
\caption{\label{fig:parameterspaceconition1} 
The parameter space (cyan) of electric BHs in which no linear instability occurs on the outer horizon $r=r_h$. 
The blue curve corresponds to the saturation condition $r-4\beta h'=0$ at $r=r_h$, and the red curve corresponds to the saturation condition of \eqref{staeven}. 
The magenta point represents the BH solution shown in the left panel of Fig.~\ref{fig1} with $(\tilde{q}_E,\tilde\beta)=(0.5,0.2)$, in which case the event horizon is at $r_h/2M=0.9117$.}
\end{figure}

In Fig.~\ref{fig:parameterspaceconition1}, we demonstrate the region of parameter spaces (cyan) for the electric BH in which all the linear stability conditions, 
i.e., \eqref{staodd}, \eqref{staeven} and \eqref{sta3}, are satisfied on the outer event horizon.
To plot this figure, we compute the BH ADM mass $M$ in the 
unit of $r_h$ for given values of $\tilde{\beta}=\beta/r_h^2$ and 
$\tilde{q}_E=q_E/(\Mpl r_h)$.
The blue and red curves represent the saturation conditions for $r-4\beta h'=0$ and \eqref{staeven} at $r=r_h$, respectively. These two curves intersect at $\beta=M^2$ when $q_E=0$. 
All the other linear stability conditions are satisfied in the cyan region. The saturation condition $r^2-8\beta(h-1)=0$ is satisfied at $\beta=-M^2/2$ when $q_E=0$, 
which coincides with the black solid curve 
in the same limit (see also Fig.~\ref{fig:parameterspace}). 
We note that the model parameters chosen in the left panel of Fig.~\ref{fig1} correspond to the magenta point in Fig.~\ref{fig:parameterspaceconition1}, which is inside the cyan region.
In summary, for the electric BH without naked singularities, the upper limit on $\beta$ is determined by the condition $r-4\beta h'>0$ at $r=r_h$.

%
\subsection{$l=0$}

For the monopole ($l=0$), the Lagrangian ${\cal L}_A$ is vanishing and hence the odd-parity perturbations 
do not propagate. 
In the even-parity sector, we choose the gauge $K=0$ and $H_1=0$ (which leaves $\Theta$ as a residual gauge degree of freedom). We also define
\be
\Phi=a_1 H_2\,.
\ee
Then, the Lagrangian (\ref{Leven}) reduces to 
\be
{\cal L}_{\rm even}=H_0 \Phi'+s_2 H_0 V
+\frac{4c_0 s_2-s_3^2 A_0'}{4a_1^2 s_2}\Phi^2
-s_1 V^2+2s_1 V \left( \delta A_0'-\dot{\delta A}_1 
\right)+\frac{s_3}{a_1}V \Phi\,, 
\label{Leven0}
\ee
where we used the relations $a_0=s_2^2/(4s_1)$, $a_3=a_1'+s_3 A_0'/2$, and $s_1=s_2/A_0'$. 

Varying ${\cal L}_{\rm even}$ with respect to $\delta A_0$ and $\delta A_1$, respectively, it follows that $(s_1 V)'=0$ and $(s_1 V)^{\cdot}=0$. 
These equations are integrated to give $V=\delta {\cal C}/s_1$, where $\delta {\cal C}$ is constant 
in space and time. 
Setting the boundary condition $\delta {\cal C}=0$ far away from the horizon\footnote{At spatial infinity, we could introduce Fourier modes for the $r$ variables. On doing this, we are left with $V=0$.}, we obtain $V=0$.
Varying Eq.~(\ref{Leven0}) with respect to $H_0$ and using $V=0$, it follows that $\Phi'=0$. 
Imposing the appropriate boundary condition at spatial infinity, we have that $\Phi=0$.
Then, the Lagrangian (\ref{Leven0}) vanishes identically.
This shows that there are no propagating degrees of freedom in the even-parity sector.

\subsection{$l=1$}

For the dipole ($l=1$), we first consider the propagation in the odd-parity sector. Since the contribution to the action arising from the field $U$ vanishes, we choose the gauge $W=0$ instead of $U=0$. Since $p_8=0=p_9$ for $l=1$, the odd-parity Lagrangian yields
\ba
{\cal L}_{\rm odd} &=& 
2 \left[ p_1 \left( Q'-\frac{2Q}{r} 
\right)^2-(p_2 \delta A+p_4 \delta A') 
\left( Q'-\frac{2Q}{r} \right)
+p_5 \dot{\delta A}^2+p_6 \delta A'^2
+p_7 \delta A^2 \right]
\nonumber \\ 
& &
-2 p_1 \left[ \dot{W}-Q'+\frac{2Q}{r} 
+\frac{1}{2p_1} \left( p_2 \delta A
+p_4 \delta A' \right)
-\chi_1 \right]^2\,.
\label{Loddl=1}
\ea
The equation of motion for $Q$ sets a constraint 
on the variable $\chi_1$, as
\be
\left( r^2 p_1 \chi_1 \right)'=0\,.
\label{chi1eq}
\ee
Strictly speaking, we cannot use this equation back into the Lagrangian, because it is a differential equation for $\chi_1$. However, we can assume an appropriate boundary condition at spatial infinity, so that Eq.~(\ref{chi1eq}) gives $\chi_1=0$. Since this is no longer a differential equation, we can substitute it into the Lagrangian. 
This leads to a reduced Lagrangian solely depending on the field $\delta A$, as
\be
{\cal L}_{\rm odd}=2p_5 \dot{\delta A}^2
-\frac{p_4^2-4p_1 p_6}{2p_1}\delta A'^2
+\frac{4 p_1^2 p_7 -p_1 (p_2^2-p_2 p_4'-p_2' p_4) 
-p_1' p_2 p_4}{2p_1^2}\delta A^2\,.
\ee
Hence there is one propagating degree of freedom $\delta A$ in the odd-parity sector. 
The ghost is absent under the condition 
\be
2p_5=\frac{r-4\beta h'}{r\sqrt{fh}}>0\,,
\ee
which is the same as the inequality (\ref{p5con}).
The radial propagation speed squared is given by
\be
c_{r,{\rm odd}}^2=\frac{1}{fh}
\frac{p_4^2-4p_1 p_6}{4p_1 p_5}=1\,,
\ee
which is luminal.
So long as $r-4\beta h'>0$, neither ghosts nor Laplacian instability arise for the field $\delta A$.

In the even-parity sector, the contributions from the two fields $K$ and $G$ appear as the combination $K-G$ and its derivatives. In this case, we choose the gauge 
\be
h_0=0\,,\qquad h_1=0\,,\qquad K=G\,.
\ee
Then, the Lagrangian ${\cal L}_{\rm even}$ in the even-parity sector is obtained by setting $h_1=0$, $h_1'=0$, $\dot{h}_1=0$, and $L=2$ in Eq.~(\ref{Leven}). 
We vary ${\cal L}_{\rm even}$ with respect to $\delta A_0$ and $\delta A_1$ and eliminate those terms from the Lagrangian.
The equation for $H_0$ is used to express the field $V$ in terms of $\chi_2$ and $\chi_2'$. 
Finally, we vary ${\cal L}_{\rm even}$ with respect to $H_1$ and solve the resulting equation for $H_1$ to remove the $H_1$-dependent term.
Then, the final reduced Lagrangian is expressed in terms of the field $\chi_2=H_2$ and its derivatives in the form 
\be
{\cal L}_{\rm even}={\cal C}_1 \dot{\cal \chi}_2'^2
-{\cal C}_2 {\cal \chi}_2''^2
+{\cal C}_3 \dot{\cal \chi}_2^2
+{\cal C}_4 {\cal \chi}_2'^2
+{\cal C}_5 \chi_2^2\,,
\label{Levenl=1}
\ee
where 
\be
{\cal C}_1=\frac{2 a_1^2 b_0 s_1^2}{s_2^2 (b_3^2-4b_0 s_6)}\,,
\qquad 
{\cal C}_2=\frac{a_1^2 s_1^2}{2 s_2^2 s_5}\,.
\ee
We do not show explicit forms of ${\cal C}_3$, ${\cal C}_4$, and ${\cal C}_5$ due to their complexities. From Eq.~(\ref{Levenl=1}), we find that there is one propagating degree of freedom $\chi_2$ in the 
even-parity sector. 
In the large frequency and momentum limits, the dominant contributions to the Lagrangian are the first two terms 
in Eq.~(\ref{Levenl=1}).
The absence of ghosts for even-parity dynamical perturbations demands that 
\be
{\cal C}_1=\frac{\Mpl^4 r^3 h^{3/2}
[r^2-8 \beta (h-1)]^2}{4q_E^2 \sqrt{f} (r-4\beta h')}>0\,.
\ee
This inequality is satisfied under the no-ghost condition $r-4\beta h'>0$ in the odd-parity sector.
The radial propagation speed squared yields
\be
c_{r,{\rm even}}^2=\frac{1}{fh} 
\frac{{\cal C}_2}{{\cal C}_1}=1\,,
\ee
which is luminal. 
Thus, for $l=1$, there are no additional linear stability conditions to those derived for $l \geq 2$.

\section{Linear stability of magnetically charged BHs}
\label{magBH}

We proceed to the analysis of the linear stability of purely magnetically charged BHs given by 
\be
q_M \neq 0\,,\qquad q_E=0\,.
\ee
In the following, we discuss the three cases: (1) $l \geq 2$, (2) $l=0$, and (3) $l=1$, separately.

\subsection{$l \geq 2$}

For $q_E=0$, we have the following relations 
\ba
& &
A_0'=0\,,\qquad 
p_2=p_4=p_{10}=p_{12}=0\,,\qquad
a_0=0\,,\qquad
b_3=b_4=0\,,\qquad
c_2=c_3=0\,,\nonumber \\
& & 
d_2=d_3=d_5=0\,,\qquad
s_2=s_3=s_4=0\,.
\ea
In this case, ${\cal L}_A$ contains the contributions of both odd- and even-parity perturbations. 
The same property also holds for ${\cal L}_B$. 
This means that, for the magnetic BH, the second-order perturbed Lagrangian does not separate into the odd- and even-parity modes. 
However, we will show that the reduced Lagrangian of dynamical perturbations can be decomposed into two sectors composed of the combinations $(\chi_1, V)$ and $(\delta A, \chi_2)$.

Varying Eq.~(\ref{L2}) with respect to $W$ and $Q$, we obtain 
\ba
W &=& \frac{2p_1 \dot{\chi}_1-p_{13}\delta A_1}{2p_8}\,,
\\
Q &=& -\frac{r^2[4 r^2 p_1^2 \chi_1' 
+2p_1 (p_3 q_M+2r^2 p_1'+4rp_1)\chi_1
- ( p_3^2 q_M -2r^2p_1 p_{11})\delta A_0 ]}
{4 r^4 p_1 p_9 - q_M^2 p_3^2}\,.
\ea
We exploit these relations to eliminate $W$, $Q$, and their derivatives from ${\cal L}_2$. 
As a next step, we vary ${\cal L}_2$ with respect to $\delta A_1$ and $\delta A_0$. 
This process leads to 
\ba
\delta A_1 &=& \frac{2L p_1 p_{13} \dot{\chi}_1
+4p_8 s_1 \dot{V}}{L(p_{13}^2-4 p_8 s_6)}\,,\\
\delta A_0 &=& [L r^2 p_1 (q_M p_3^2 -2r^2 p_1 p_{11} ) 
\chi_1' + Lr \{ 
p_1 p_3 ( 2q_M p_3+2 r^3 p_9 - rq_M p_{11} ) 
-4r^2 p_1^2 p_{11} 
+ r q_M  p_1' p_3^2 - 2r^3 p_1 p_1' p_{11}\} \chi_1 
\nonumber \\
& & +(q_M^2 p_3^2-4 r^4 p_1 p_9)
(s_1V'+s_1'V)]/[L\{ r^4p_1 p_{11}^2
-r^2q_M  p_3^2 p_{11}-4r^4 p_1 p_9 s_5 
+ p_3^2 (r^4 p_9+ q_M^2 s_5 )\}]\,,
\ea
which are used to remove $\delta A_1$, $\delta A_0$, and their derivatives from ${\cal L}_2$. 

The next procedure is to introduce the dynamical field $\chi_2=H_2-Lh_1/r$ as in Eq.~\eqref{chi2} and then express $H_2$ in terms of $\chi_2$ and $h_1$. 
Varying ${\cal L}_2$ with respect to $H_0$ and using the relations among coefficients presented in Appendix A, we can solve the resulting equation for $h_1$, as 
\be
h_1=-\frac{r[2r h a_1\chi_2' 
+(2 rh a_1'+ La_1) \chi_2 + 2L rh (a_6 \delta A'
+a_7 \delta A)]}{L[La_1 + 2h( r^2 a_5 + ra_1' 
-a_1)]}\,.
\ee
This equation is used to eliminate $h_1$ and its derivatives from ${\cal L}_2$. 
The final process is to vary ${\cal L}_2$ with 
respect to $H_1$, giving 
\ba
H_1 &=&
-r[ 2 L rh (a_1'a_6+r a_5a_6-a_1 a_7)\dot{\delta A}
+2rh a_1 (a_1'+2r a_5)\dot{\chi}_2
+L a_1 a_6 (L \dot{\delta A}-2h r\dot{\delta A}'
-2h\dot{\delta A}) \nonumber\\
& &+a_1^2 (L \dot{\chi}_2-4h \dot{\chi}_2
-2rh \dot{\chi}_2') ]
/[L a_1 \{ a_1(L-2h) + 2rh (a_1'+r a_5) \}]\,.
\ea
On using this relation, the second-order action can be expressed in the form 
\be
{\cal L}_2= 
\dot{\vec{\mathcal{Z}}}^{t} {\bm K}
\dot{\vec{\mathcal{Z}}}
+\vec{\mathcal{Z}}'^{t} {\bm G}\vec{\mathcal{Z}}'
+\vec{\mathcal{Z}}^{t} {\bm M}\vec{\mathcal{Z}}
+\vec{\mathcal{Z}}'^{t} {\bm S} \vec{\mathcal{Z}}\,,
\label{LtotalD}
\ee
where ${\bm K},{\bm G},{\bm M}$ are $4 \times 4$ symmetric matrices, ${\bm S}$ is a $4 \times 4$ antisymmetric matrix, and 
\be
\vec{\mathcal{Z}}^t= \left( \chi_1, \delta A, \chi_2, V \right)\,,
\ee
The nonvanishing elements of ${\bm K},{\bm G},{\bm M},{\bm S}$ are the (11), (22), (33), (44), (14), (41), (23), (32) components. 
This means that the dynamics of four dynamical perturbations separates into the following two sectors:
\be
{\rm (I)}~
\vec{\mathcal{Z}}_1^t= \left( \chi_1,  V \right)\,, 
\qquad 
{\rm (II)}~
\vec{\mathcal{Z}}_2^t= \left( \delta A,  \chi_2 \right)\,. 
\ee
The sector (I) is composed of odd-parity gravitational perturbation $\chi_1$ and even-parity vector-field perturbation $V$, whereas the sector (II) consists of odd-parity vector-field perturbation $\delta A$ and even-parity gravitational perturbation $\chi_2$. Similar splitting also arises for perturbations of the magnetic BHs in Einstein-Maxwell theory with corrections from nonlinear electrodynamics \cite{Nomura:2020tpc, Nomura:2021efi}.
In the following, we derive the linear stability conditions of BHs by separating the reduced Lagrangian (\ref{LtotalD}) into two sectors.

\subsubsection{No-ghost conditions}

In the sector (I), the no-ghost conditions for the perturbations $\chi_1$ and $V$ are given by 
\ba
& & 
K_{11}=\frac{L \sqrt{h} (\Mpl^2 r^4+
4\beta q_M^2)^2}{4(L-2)\Mpl^2 r^6 f^{3/2}}>0\,,
\label{nogoqM1}\\
& &
K_{11}K_{44}-K_{14}^2
=\frac{h [r^2-8 \beta(h-1)]^2(\Mpl^2 r^4+
4\beta q_M^2)^3}{8(L-2) \Mpl^2 r^6 f^3
[\Mpl^2 r^4+4\beta \Mpl^2 (h-1)r^2
+6\beta q_M^2]}>0\,.
\label{nogoqM2}
\ea
For the multipole $l \geq 2$, the first inequality (\ref{nogoqM1}) is automatically satisfied. 
In the sector (II), the absence of ghosts for the perturbations $\chi_2$ and $\delta A$ demands that 
\ba
& & 
K_{33}
=\frac{4h^{3/2}(\Mpl^2 r^4+
4\beta q_M^2) [\Mpl^4 (L - 2) r^8 
+2\Mpl^2 q_M^2 r^6 + 4 \Mpl^2 \beta q_M^2
(L + 2h - 4) r^4 + 12 \beta q_M^4 r^2 
- 768 \beta^2 q_M^4 h]}
{\sqrt{f}L r^2[ 2\Mpl^2 (L - 3h + 1) r^4 
- q_M^2 r^2 + 8 \beta q_M^2 (L - 8 h)]^2}>0,
\label{nogoqM3}\nonumber \\ \\
& &
K_{22}K_{33}-K_{23}^2
=\frac{2h\Mpl^2 (L-2)(\Mpl^2 r^4+
4\beta q_M^2)[\Mpl^2 r^6 +4 \beta \Mpl^2 (h - 1)r^4 
+6 \beta q_M^2 r^2-384 \beta^2 q_M^2 h]}
{f[2\Mpl^2 (L - 3 h + 1) r^4 
-q_M^2 r^2 + 8\beta q_M^2 (L - 8h)]^2}>0\,.
\label{nogoqM4}
\ea
The conditions (\ref{nogoqM1})-(\ref{nogoqM4}) are satisfied under the following inequalities
\ba
{\cal F}_1 &\equiv& \Mpl^2 r^4+4\beta q_M^2>0\,,\label{k1}\\
{\cal F}_2 &\equiv& \Mpl^2 r^6 +4 \beta \Mpl^2 (h - 1)r^4 
+6 \beta q_M^2 r^2-384 \beta^2 q_M^2 h>0\,,\label{k2}\\
{\cal F}_3 &\equiv& 
\Mpl^4 (L - 2) r^8 
+2\Mpl^2 q_M^2 r^6 + 4 \Mpl^2 \beta q_M^2
(L + 2h - 4) r^4 + 12 \beta q_M^4 r^2 
- 768 \beta^2 h q_M^4>0\,.\label{k3}
\ea
For small $\beta$ close to 0, all of them are consistently satisfied. 

\subsubsection{Radial propagation speeds}

To derive the radial propagation speeds $c_{r}$ in the sector (I), we solve the following equation 
\be
{\rm det} \left( fh c_r^2 {\bm K}_{\rm I}
+{\bm G}_{\rm I} \right)=0\,,
\ee
where 
\be
{\bm K}_{\rm I}=
\begin{pmatrix}
   K_{11} & K_{14} \\
   K_{14} & K_{44}
\end{pmatrix}\,,\qquad 
{\bm G}_{\rm I}=
\begin{pmatrix}
   G_{11} & G_{14} \\
   G_{14} & G_{44}
\end{pmatrix}\,.
\label{KGI}
\ee
Then, we obtain the two solutions
\ba
& &
c_{r1}^2=1\,,
\label{cr1}\\
& &
c_{r2}^2=\frac{\Mpl^2 r^6+4 \Mpl^2 \beta (h - 1) r^4 
+6 \beta q_M^2 r^2 }{\Mpl^2 r^6+4 \Mpl^2 
\beta (h - 1) r^4 
+6 \beta q_M^2 r^2 - 384 \beta^2 q_M^2 h}\,.
\label{cr2}
\ea
For the other sector (II), the radial propagation speeds are known by solving 
\be
{\rm det} \left( fh c_r^2 {\bm K}_{\rm II}
+{\bm G}_{\rm II} \right)=0\,,
\ee
where 
\be
{\bm K}_{\rm II}=
\begin{pmatrix}
   K_{22} & K_{23} \\
   K_{23} & K_{33}
\end{pmatrix}\,,\qquad 
{\bm G}_{\rm II}=
\begin{pmatrix}
   G_{22} & G_{23} \\
   G_{23} & G_{33}
\end{pmatrix}\,.
\ee
The resulting solutions are given by 
\ba
& &
c_{r3}^2=1\,,\\
& &
c_{r4}^2=\frac{\Mpl^2 r^6+4 \Mpl^2 \beta (h - 1) r^4 
+6 \beta q_M^2 r^2}{\Mpl^2 r^6+4 \Mpl^2 
\beta (h - 1) r^4 
+6 \beta q_M^2 r^2 - 384 \beta^2 q_M^2 h}\,,
\label{cr4}
\ea
and hence $c_{r1}^2=c_{r3}^2$ and $c_{r2}^2=c_{r4}^2$. 
Thus, the two propagation speeds in the two sectors (I) and (II) coincide with each other. 
The difference from the electric BH is that the two squared propagation speeds $c_{r2}^2$ and $c_{r4}^2$ deviate from 1, while the other two are 1. Under the inequality (\ref{k2}), we have $\Mpl^2 r^6+4 \Mpl^2 \beta (h - 1) r^4+6 \beta q_M^2 r^2>0$ outside the horizon ($h>0$) and hence the radial Laplacian stability conditions $c_{r2}^2>0$ and $c_{r4}^2>0$ are satisfied. We note that, on the outer 
horizon ($h \to 0$) and at spatial infinity, 
both $c_{r2}^2$ and $c_{r4}^2$ approach 1. 
In the limit that $\beta \to 0$, we also have
$c_{r2}^2 \to 1$ and $c_{r4}^2 \to 1$.
\subsubsection{Angular propagation speeds}

For the discussion of the angular propagation in the limit $L \gg 1$, we use the fact that the dominant contributions to the perturbation equations arise from the two matrices ${\bm K}$ and ${\bm M}$ in Eq.~(\ref{LtotalD}), while the contribution from 
${\bm S}$ can be neglected.
In the sector (I), the angular propagation speeds $c_{\Omega}$ are known by solving 
\be
{\rm det} \left( fL c_{\Omega}^2 
{\bm K}_{\rm I}+r^2 {\bm M}_{\rm I} 
\right)=0\,,
\ee
where ${\bm K}_{\rm I}$ is given in Eq.~(\ref{KGI}), and 
\be
{\bm M}_{\rm I}=
\begin{pmatrix}
   M_{11} & M_{14} \\
   M_{14} & M_{44}
\end{pmatrix}\,.
\ee
Then, we obtain the two solutions
\ba
c_{\Omega 1}^2 
&=& \frac{\Mpl^2 r^6+4 \Mpl^2 \beta (h - 1)r^4 
+6 \beta q_M^2 r^2}
{(\Mpl^2 r^4+4\beta q_M^2)
[r^2-8 \beta (h-1)]}\,,\\
c_{\Omega 2}^2 
&=& \frac{\Mpl^2 r^4 [\Mpl^2 r^6+4 \Mpl^2 \beta 
(h - 1)r^4 +6 \beta q_M^2 r^2-96\beta^2 q_M^2 h]}{(\Mpl^2 r^4+4\beta q_M^2)
[\Mpl^2 r^6+4 \Mpl^2 \beta (h - 1)r^4 
+6 \beta q_M^2 r^2-384\beta^2 q_M^2 h]}\,.
\ea
For the other sector (II), we solve the following equation 
\be
{\rm det} \left( fL c_{\Omega}^2 
{\bm K}_{\rm II}+r^2 {\bm M}_{\rm II} 
\right)=0\,,
\ee
where 
\be
{\bm M}_{\rm II}=
\begin{pmatrix}
   M_{22} & M_{23} \\
   M_{23} & M_{33}
\end{pmatrix}\,.
\ee
This gives rise to the following solutions 
\ba
c_{\Omega 3}^2 &=& \frac{\Mpl^2 r^4}{\Mpl^2 r^4
+4\beta q_M^2}\,,\\
c_{\Omega 4}^2 &=& \frac{\Mpl^2 r^4
[\Mpl^2 r^6-8 \Mpl^2 \beta (h - 1)r^4 
+96 \beta^2 q_M^2 h]}{(\Mpl^2 r^4+4\beta q_M^2)
[\Mpl^2 r^6+4 \Mpl^2 \beta (h - 1)r^4 
+6 \beta q_M^2 r^2-384\beta^2 q_M^2 h]}\,.
\ea
Under the inequalities (\ref{k1}) and (\ref{k2}), both $c_{\Omega 2}^2$ and $c_{\Omega 3}^2$ are positive. To avoid the Laplacian instabilities associated with $c_{\Omega 1}^2$ and $c_{\Omega 4}^2$, we require that 
\ba
{\cal F}_4 &\equiv& r^2-8 \beta (h-1)>0\,,
\label{k4} \\
{\cal F}_5 &\equiv& \Mpl^2 r^6-8 \Mpl^2 
\beta (h - 1)r^4 +96 \beta^2 q_M^2 h>0\,.
\label{k5} 
\ea
Thus, there are neither ghosts nor Laplacian instabilities under the conditions (\ref{k1}), (\ref{k2}), (\ref{k3}), (\ref{k4}), and (\ref{k5}). 
In the limit that $\beta \to 0$, all these conditions are trivially satisfied, with the luminal radial and angular propagation speeds. 
We note that, for $\beta \neq 0$, the four squared angular propagation speeds $c_{\Omega 1}^2$, $c_{\Omega 2}^2$, $c_{\Omega 3}^2$, and $c_{\Omega 4}^2$ are different from each other. This suggests the breaking of the isospectrality of quasinormal modes between the sectors (I) and (II). 
Moreover, the eikonal correspondence is also broken as in the case of electric BHs.

\begin{figure}[ht]
\begin{center}
\includegraphics[height=2.8in,width=4.5in]{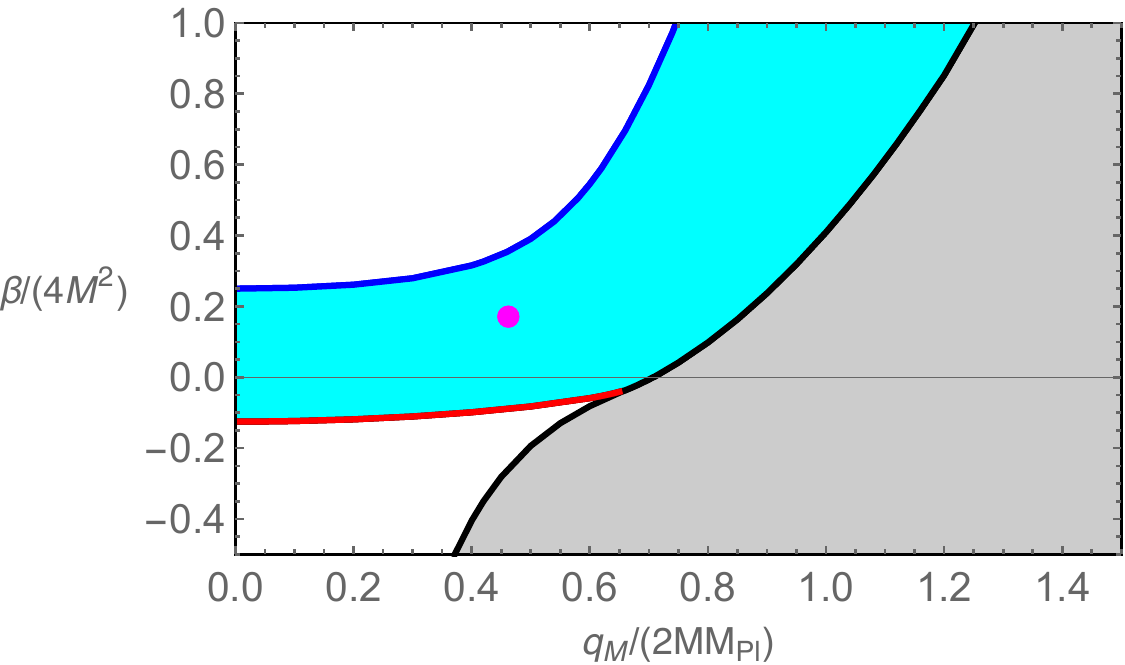}
\end{center}
\caption{\label{fig:parameterspaceconition2} 
The parameter space (cyan) of magnetic BHs in which no instability occurs on the outer horizon. 
The blue and red curves represent the saturation conditions for $\mathcal{F}_2=0$ and $\mathcal{F}_4=0$ on the horizon, respectively. 
The magenta point represents the BH solution shown in the right panel of Fig.~\ref{fig1} with $(\tilde{q}_M,\tilde\beta)=(0.5,0.2)$, in which case the outer event horizon is at $r_h/2M=0.9266$.}
\end{figure}

In Fig.~\ref{fig:parameterspaceconition2}, we demonstrate the region of the parameter space (cyan) of magnetic BHs in which all the stability conditions (\ref{k1})-(\ref{k3}) and (\ref{k4})-(\ref{k5}) are satisfied on the outermost event horizon. 
Since $h=0$ at $r=r_h$, the saturation conditions $\mathcal{F}_4=0$ and $\mathcal{F}_5=0$ are equivalent 
to each other. Moreover, for $l \geq 2$, the inequality
(\ref{k3}) does not give a tighter bound than 
the condition (\ref{k2}). 
The blue and red curves represent the saturation conditions $\mathcal{F}_2=0$ and $\mathcal{F}_4=0$ on the horizon, which intersect $\beta=M^2$ and $\beta=-M^2/2$ when $q_M=0$, respectively. 
Note that the inequality (\ref{k1}) does not give
an additional bound to those obtained by the 
conditions (\ref{k2}) and (\ref{k4}).
The magenta point in Fig.~\ref{fig:parameterspaceconition2} represents the model parameters chosen in the right panel of Fig.~\ref{fig1}, which is within the cyan region. 
In summary, the upper limit on $\beta$ under which the magnetic BHs are linearly stable is determined by the condition $\mathcal{F}_2>0$. For the magnetic BHs without naked singularities, $\beta$ is bounded from below by the other condition $\mathcal{F}_4>0$.

\subsection{$l=0$}

For $l=0$, all the terms associated with the odd-parity perturbations vanish in the total second-order Lagrangian and hence the odd-modes do not propagate.
In the even-parity sector, we choose the gauge conditions $K=0, H_1=0$ and introduce the field $\Phi=a_1 H_2$. Then, the Lagrangian of even-parity perturbations yields 
\be
{\cal L}_{\rm even}=H_0 \Phi'+\frac{c_0}{a_1^2}\Phi^2
-s_1 V^2+2s_1 V \left( \delta A_0'-\dot{\delta A}_1 
\right)\,.
\label{Leven0d}
\ee
Varying this with respect to $\delta A_0$, $\delta A_1$, $V$ and choosing appropriate boundary conditions at spatial infinity, we have that $V=0$ and $\Phi=0$.
Then, the Lagrangian (\ref{Leven0d}) vanishes and hence the even modes do not propagate either.

\subsection{$l=1$}

For $l=1$, analogous to the discussion of 
electric BHs, we choose the following gauge condition 
\be
W=0\,,\qquad h_0=0\,,\qquad h_1=0\,,\qquad 
K=G\,.
\ee
The total Lagrangian ${\cal L}_2$ can be obtained from Eq.~(\ref{L2}) by setting $W=0$, $\dot{W}=0$, $h_1=0$, $\dot{h}_1=0$, $h_1'=0$, and $L=2$ in Eqs.~(\ref{LA}) 
and (\ref{LB}). 
We first eliminate the field $Q$ by using its equation of motion. As a next step, we vary ${\cal L}_2$ with respect to $\delta A_0$. On using the relations 
\be
p_9=\frac{r^4 p_1 p_{11}^2
-r^2 q_M p_3^2 p_{11}+q_M^2 p_3^2 s_5}
{r^4 (4p_1 s_5-p_3^2)}\,,\qquad 
p_{11}=\frac{2q_M}{r^2}s_5\,,
\ee
we find that the equation of motion for $\delta A_0$ gives a constraint on other fields as $(q_M s_1 V+2r^2 p_1 \chi_1)'=0$. 
Choosing an appropriate boundary condition at spatial infinity, we have that 
\be
V=-\frac{2r^2 p_1}{q_M s_1}\chi_1\,,
\ee
which will be used to remove the field 
$V$ from ${\cal L}_2$. 
Varying ${\cal L}_2$ with respect to $\delta A_1$ and $H_1$, respectively, we obtain
\be
\delta A_1=\frac{4r^2 p_1 p_8}{q_M p_{13}^2} 
\dot{\chi}_1\,,\qquad 
H_1=-\frac{b_1 \dot{\chi}_2+2b_5 \dot{\delta A}}{4b_0}\,,
\ee
so that these fields are eliminated from the Lagrangian.
Finally, variation of ${\cal L}_2$ with 
respect to $H_0$ gives 
\be
a_1 \chi_2'+2a_6 \delta A'
+(a_1'+2a_4)\chi_2
+2a_7 \delta A=0\,,
\label{H0re}
\ee
where $\chi_2=H_2$.
We introduce the new field
\be
\chi=a_1 \chi_2+2a_6 \delta A\,.
\ee
On using Eq.~(\ref{H0re}), we can express $\chi_2$ and $\delta A$ by using $\chi'$ and $\chi$.
After eliminating the fields $\chi_2$ and $\delta A$, the final Lagrangian is given by ${\cal L}_2={\cal L}_{\chi_1}+{\cal L}_{\chi}$, where 
\ba
{\cal L}_{\chi_1} &=& {\cal K}_1 \dot{\chi}_1^2 
-{\cal G}_1 \chi_1'^2+{\cal M}_1 \chi_1^2\,,
\label{Lchi1} \\
{\cal L}_{\chi} &=& {\cal K} \dot{\chi}'^2
-{\cal G} \chi''^2+ \tilde{{\cal K}} \dot{\chi}^2
+ \tilde{{\cal G}} \chi'^2
+\tilde{{\cal M}} \chi^2\,,
\label{Lchi} 
\ea
where ${\cal K}_1$, ${\cal K}$, etc.\ are functions of $r$. 
This means that the dipole perturbations have two dynamical propagating degrees of freedom $\chi_1$ and $\chi$.

From the Lagrangian (\ref{Lchi1}), the field $\chi_1$ has neither ghosts nor radial Laplacian instability under the two conditions
\ba
{\cal K}_1 &=& \frac{\sqrt{h} (\Mpl^2 r^4+
4 \beta q_M^2)^3}{4q_M^2 r^4 f^{3/2} 
[\Mpl^2 r^4+4\Mpl^2 \beta r^2 (h-1)
+6 \beta q_M^2]}>0\,,\label{K1c}\\
c_{r1,{\rm even}}^2 &=&
\frac{1}{fh}
\frac{{\cal G}_1}{{\cal K}_1}
=\frac{\Mpl^2 r^6+4 \Mpl^2 \beta (h - 1) r^4 
+6 \beta q_M^2 r^2}{\Mpl^2 r^6+4 \Mpl^2 
\beta (h - 1) r^4 
+6 \beta q_M^2 r^2 - 384 \beta^2 q_M^2 h}>0\,.
\ea
Similarly, from the Lagrangian (\ref{Lchi}), the absence of ghosts and radial Laplacian instability for the field $\chi$ requires that 
\ba
{\cal K} &=& \frac{r^6 \sqrt{h} 
(\Mpl^2 r^4+4 \beta q_M^2)
[\Mpl^2 r^6+4 \Mpl^2 \beta (h - 1) r^4 
+6 \beta q_M^2 r^2-384 \beta^2 q_M^2 h]}
{q_M^2 f^{3/2}[\Mpl^2 r^6 + 40 \Mpl^2   
\beta (h - 1)r^4+ 12 \beta q_M^2 r^2 
- 96 \beta^2 q_M^2]^2}>0\,,\label{Kc}\\
c_{r2,{\rm even}}^2 &=&
\frac{1}{fh}
\frac{{\cal G}}{{\cal K}}
=\frac{\Mpl^2 r^6+4 \Mpl^2 \beta (h - 1) r^4 
+6 \beta q_M^2 r^2}{\Mpl^2 r^6+4 \Mpl^2 
\beta (h - 1) r^4 
+6 \beta q_M^2 r^2 - 384 \beta^2 q_M^2 h}>0\,.
\ea
Thus, we find that $c_{r1,{\rm even}}^2$ and $c_{r2,{\rm even}}^2$ are equivalent to each other and that they are the same as Eqs.~(\ref{cr2}) and (\ref{cr4}) derived for $l \geq 2$. 
Moreover, under the inequalities (\ref{k1})-(\ref{k3}), the no-ghost conditions (\ref{K1c}) and (\ref{Kc}) are satisfied. 
Hence the dipole perturbations\footnote{An equivalent result is found by setting the gauge $H_2=0$, $K=G$, $h_0=0$, and $W=0$, which validates the results 
and assumptions on the gauge choice.} do not provide additional linear stability conditions to those obtained for $l \geq 2$.

\section{Conclusions}
\label{consec}

In this paper, we studied the BH perturbations on the static and spherically symmetric background (\ref{metric_bg}) in Einstein-Maxwell-HVT theory given by the action (\ref{action}). 
The HVT interaction ${\cal L}_{\rm HVT}=\beta L^{\mu \nu \alpha \beta} F_{\mu \nu}F_{\alpha \beta}$ is a unique combination of the Lagrangian in $U(1)$ gauge-invariant vector-tensor theories keeping the equations of motion up to second order.
There are electrically charged or/and magnetically charged BH solutions where the RN geometry is modified by the coupling constant $\beta$. 
The HVT interaction breaks the electric-magnetic duality at the background level, such that the metric components $f$ and $h$ for the electric BH are different from those for the magnetic BH with the same charge. 
In Fig.~\ref{fig:parameterspace}, we clarified the parameter spaces in which the event horizons exist without naked singularities for both electric and magnetic BHs.

In Sec.~\ref{BHpersec}, we performed the general formulation of BH perturbations in Einstein-Maxwell-HVT theory that are valid for dyon BHs with mixed electric and magnetic charges. Choosing the gauge (\ref{RWgauge}) for the multipoles $l \geq 2$, the total second-order Lagrangian ${\cal L}$ consists of two contributions 
${\cal L}_A$ and ${\cal L}_B$, see Eqs.~(\ref{LA}) 
and (\ref{LB}). 
Introducing Lagrange multipliers $\chi_1$ and $V$, which are given by Eqs.~(\ref{chi1}) and (\ref{V}) respectively, the reduced Lagrangian can be expressed in terms of the four dynamical fields $\chi_1$, $\delta A$, $\chi_2$, 
$V$ and their derivatives. 
These fields are associated with the gravitational and vector-field perturbations in the odd- and even-parity sectors. Thus, the HVT coupling does not increase the number of propagating DOFs in comparison to those in Einstein-Maxwell theory. 

In Sec.~\ref{eleBH}, we studied the linear stability of purely electrically charged BHs by separating the analysis depending on the multipoles $l$.
For $l \geq 2$, the second-order Lagrangian can be decomposed into those of the odd- and even-parity sectors containing the combinations of dynamical perturbations $(\chi_1, \delta A)$ and 
$(\chi_2, V)$. We found that, in the high frequency and momentum limit, all of the four dynamical propagation speeds along the radial direction are equivalent to 1.
In the large $l$ limit, the four angular propagation speeds are different from each other, with their deviations from 1 weighed by the coupling $\beta$. 
This shows the breaking of eikonal correspondence between the peak of at least one of the potentials of dynamical perturbations and the radius of photon sphere. In Fig.~\ref{fig:parameterspaceconition1}, we plotted the parameter space of $(\beta, q_E)$ in which the BHs without naked singularities are prone to neither ghost nor Laplacian instabilities (cyan). 
For $l=0$, there are no dynamical perturbations propagating on the static and spherically symmetric 
background. 
For $l=1$, two perturbations $\delta A$ and $\chi_2$ propagate, but there are no additional linear stability conditions than those derived for $l \geq 2$.

In Sec.~\ref{magBH}, we applied the linear perturbation theory to purely magnetically charged BHs.
For $l \geq 2$, we showed that the dynamics of four dynamical perturbations can be decomposed into the two sectors: (I) $(\chi_1, V)$ and (II) $(\delta A, \chi_2)$. 
From each sector, we obtained the same radial propagation speeds given by Eqs.~(\ref{cr1}) and~(\ref{cr2}), one of which deviates from 1 by the nonvanishing coupling $\beta$. 
On the other hand, the angular propagation speeds of four dynamical perturbations are different from each other. Hence the eikonal correspondence is also broken for the magnetic BH. 
Given the same amount of charges, the linear stability conditions (absence of ghosts and Laplacian instabilities) for the magnetic BH differ from those for the electric BH. 
This is the manifestation of breaking of electric-magnetic duality at the level of linear perturbations. 
In Fig.~\ref{fig:parameterspaceconition2}, we presented the parameter space of $(\beta, q_M)$ in which the magnetic BH without naked singularities is linearly stable. 
For $l=0$, we showed that no dynamical perturbations propagate. For $l=1$, there are two dynamical perturbations, but their linear stability does not add new conditions to those obtained for $l \geq 2$.

We thus showed that the HVT coupling gives rise to nontrivial BH solutions whose properties are different between the electrically and magnetically charged cases  both at the levels of background and perturbations.
It will be of interest to compute the quasinormal modes 
of those BHs to see whether the isospectrality of the two sectors is broken. 
To study the light-ray bending induced by the coupling $\beta$ will be also intriguing, especially in connection to the observations of BH shadows.

\section*{Acknowledgements}

We thank the organizers of the Gravity and Cosmology 2024 workshop held at YITP, Kyoto University, during which this work was initiated. 
CYC is supported by the Special Postdoctoral Researcher (SPDR) Program at RIKEN. The work of ADF was supported by the Japan Society for the Promotion of Science Grants-in-Aid for Scientific Research No.~20K03969. 
ST was supported by the Grant-in-Aid for Scientific 
Research Fund of the JSPS No.~22K03642 and Waseda University 
Special Research Project No.~2023C-473.

\section*{Appendix~A:~Coefficients in the second-order action}
\label{AppA}

The $r$-dependent coefficients in Eqs.~(\ref{LA}) 
and (\ref{LB}) are 
\ba
& &
p_1=-\frac{a_1}{2rf}\,,\qquad 
p_2=-\frac{q_E (r^2-8 \beta h)}
{r^2[r^2 -8 \beta (h-1)]}\,,\qquad p_3=\frac{a_6}{f}\,,\qquad
p_4=-\frac{4\beta h^{3/2}}{r\sqrt{f}}A_0'\,,\nonumber \\
& &
p_5=s_5\,,\qquad p_6=s_6\,,\qquad
p_7=-\frac{L[2\beta \{ h f'^2-f(2f''h+f'h')\}+f^2]}
{2r^2 f^{3/2} \sqrt{h}}\,,\nonumber \\
& &
p_8=-\sqrt{h}[\Mpl^2 (\Mpl^2 f-4 \beta h A_0'^2) (L-2)r^6 
+ 2q_M^2 ( \Mpl^2 f+2 \beta h A_0'^2) r^4 
- 4 q_M^2 \{ 4 hA_0'^2 (L + 6h - 4) \beta 
\nonumber \\
& &\qquad\, 
-\Mpl^2 f (L + 2 h - 4)  \} \beta r^2
+12 \beta f q_M^4 ]/[4r^4 \sqrt{f} (\Mpl^2 r^4 
+ 4\beta q_M^2)]\,,\nonumber \\
& &
p_9=[\Mpl^2 (\Mpl^2 f-4 \beta h A_0'^2) (L-2)r^8 
+ 2q_M^2 (\Mpl^2 f+2 \beta h A_0'^2 ) r^6 
+4 \beta q_M^2  (\Mpl^2 f-4\beta h A_0'^2)
(L + 2h - 4) r^4
\nonumber \\
& &\qquad 
+12 \beta f q_M^4 (r^2-16 \beta h)]/[4r^6 f^{3/2} \sqrt{h} 
(\Mpl^2 r^4 + 4\beta q_M^2)]\,,\nonumber \\
& &
p_{10}=-\frac{16(L-2)\sqrt{fh}\beta q_E}
{r[r^2-8 \beta (h-1)]^2}s_5\,,\qquad
p_{11}=\frac{2q_M}{r^2}s_5\,,\qquad
p_{12}=-\frac{q_E q_M (r^2-8 \beta h)}
{r^4[r^2 -8 \beta (h-1)]}\,,
\qquad 
p_{13}=\frac{2q_M}{r^2}s_6\,,\nonumber \\
& &
a_0=\frac{q_E}{8}A_0'\,,\qquad 
a_1=-\frac{\sqrt{fh} (\Mpl^2 r^4+4 \beta q_M^2)}{2r^3}\,,
\qquad 
a_2=-\frac{a_1}{r}\,,\qquad 
a_3=a_1'+\frac{1}{2}s_3 A_0'\,,\nonumber \\
& &
a_4=\frac{a_1}{2rh}\,,\qquad 
a_5=\frac{\sqrt{f}}{4r^4 \sqrt{h}} 
\left[ \Mpl^2 r^3 (rh'+2h)+\beta \left( 
4q_M^2 h'-\frac{8r^3 h^2}{f}A_0'^2
\right)\right]\,,\nonumber \\
& &
a_6=\frac{12 \sqrt{fh}\beta q_M}{r^3}\,,\qquad 
a_7=\frac{q_M \sqrt{f}[r^2+8 \beta (rh'-6h)]}{2r^4 \sqrt{h}}\,,
\nonumber \\
& &
b_0=-\frac{a_1}{2rf}\,,\qquad 
b_1=-\frac{2}{f}a_1\,,\qquad 
b_2=\frac{a_1}{rf}\,,\qquad 
b_3=\frac{4\beta h^{3/2}}{r \sqrt{f}}A_0'\,,\qquad 
b_4=\frac{q_M}{r^2}b_3\,,\qquad 
b_5=-\frac{a_6}{f}\,,\nonumber \\
& &
c_0=-\frac{1}{2} a_3
-\frac{2\beta h^{3/2}}{\sqrt{f}}A_0'^2\,,\qquad
c_1=-\frac{\sqrt{h}[\Mpl^2 r^3 (rf' + 2f) 
+ 4\beta ( q_M^2 f'-6r^3 h A_0'^2)]}{4r^4 \sqrt{f}}\,,
\nonumber \\
& &
c_2=-\frac{b_3}{h}\,,\qquad 
c_3=\frac{q_M}{r^2}c_2\,,\qquad 
c_4=\frac{q_M[4 \beta h \{ r^2+8\beta (1 - 3h)\}A_0'^2
-\Mpl^2f \{ r^2 - 8\beta (h - 1)\}]}
{2\sqrt{fh} (\Mpl^2 r^4+4 \beta q_M^2)}\,,\nonumber \\
& &
d_0=-\frac{a_1}{2rf}\,,\qquad
d_1=\frac{\sqrt{h}[q_M^2 (16 \beta f'h - rf) 
+ 2 \Mpl^2 r^3 h (rf' + f) + 
r^3 h (r^2-24 \beta h )A_0'^2]}
{4r^5 \sqrt{f}}\,,\nonumber \\
& &
d_2=b_3\,,\qquad 
d_3=\frac{q_M}{r^2}b_3\,,\qquad 
d_4=\frac{a_6}{f}\,,\qquad 
d_5=p_2\,,\qquad 
d_6=\frac{2q_M}{r^2}s_6\,,\nonumber \\
& &
s_1=\frac{\sqrt{h}[r^2-8\beta (h-1)]}{2\sqrt{f}}
=\frac{q_E}{2A_0'}\,,\qquad 
s_2=\frac{q_E}{2}\,,\qquad 
s_3=-\frac{r^2+8 \beta (1-3h)}{2[r^2-8 \beta (h-1)]}q_E\,,\qquad
s_4=-2b_3\,,\nonumber \\
& &
s_5=\frac{r-4\beta h'}{2r\sqrt{fh}}\,,\qquad
s_6=-\frac{\sqrt{h} (rf-4\beta f' h)}{2r \sqrt{f}}\,,
\ea
where $A_0'$ is given by Eq.~(\ref{back4}).

\section*{Appendix~B:~Coefficients in the 
equations of motion for $\chi_2$ and $V$}
\label{AppB}

In Eqs.~(\ref{perre1}) and (\ref{perre2}), the coefficients 
are given by 
\ba
& &
\mu_1=\frac{\tilde{K}_{12}(\tilde{G}_{12}' - \tilde{S}_{12}) 
- \tilde{K}_{22}\tilde{G}_{11}'}
{\tilde{K}_{11} \tilde{K}_{22}-\tilde{K}_{12}^2}\,,
\qquad 
\mu_2=\frac{2 \tilde{K}_{22} \tilde{M}_{11}
-\tilde{K}_{12} (\tilde{S}_{12}'+2 \tilde{M}_{12})}
{2(\tilde{K}_{11} \tilde{K}_{22}-\tilde{K}_{12}^2)}\,,
\nonumber \\
& &
\mu_3=\frac{\tilde{K}_{12} \tilde{G}_{22}'
-\tilde{K}_{22}(G_{12}'+ \tilde{S}_{12})}
{\tilde{K}_{11} \tilde{K}_{22}-\tilde{K}_{12}^2}\,,
\qquad 
\mu_4=\frac{\tilde{K}_{22} (2 \tilde{M}_{12} 
-\tilde{S}_{12}')
-2 \tilde{K}_{12} \tilde{M}_{22}}{2(\tilde{K}_{11} \tilde{K}_{22}-\tilde{K}_{12}^2)}\,,\\
& &
\nu_1=\frac{\tilde{K}_{12}(\tilde{G}_{12}' 
+ \tilde{S}_{12})-\tilde{K}_{11}\tilde{G}_{22}'}
{\tilde{K}_{11} \tilde{K}_{22}-\tilde{K}_{12}^2}\,,\qquad
\nu_2=\frac{2 \tilde{K}_{11}\tilde{M}_{22}
+\tilde{K}_{12}(\tilde{S}_{12}'-2\tilde{M}_{12})}
{2(\tilde{K}_{11} \tilde{K}_{22}-\tilde{K}_{12}^2)}\,,\nonumber\\
& &
\nu_3=\frac{\tilde{K}_{12}\tilde{G}_{11}'
+\tilde{K}_{11}(\tilde{S}_{12}-\tilde{G}_{12}')}
{\tilde{K}_{11} \tilde{K}_{22}-\tilde{K}_{12}^2}\,,\qquad 
\nu_4=\frac{\tilde{K}_{11}(\tilde{S}_{12}'+2\tilde{M}_{12})
-2\tilde{K}_{12}\tilde{M}_{11}}{2(\tilde{K}_{11} \tilde{K}_{22}-\tilde{K}_{12}^2)}\,.
\ea

\bibliographystyle{mybibstyle}
\bibliography{bib}

\end{document}